\pdfoutput=1
\documentclass[11pt]{article}

\usepackage[a4paper,margin=2.5cm]{geometry}
\usepackage{graphicx}
\usepackage{amsmath,amssymb}
\usepackage{booktabs}
\usepackage{textcomp}
\usepackage[superscript]{cite}                    
\usepackage[colorlinks=true,linkcolor=blue,citecolor=blue,urlcolor=blue]{hyperref}
\usepackage{tabularx}

\graphicspath{{figures/}}

\newcommand{\csd}{Claude-SD}
\newcommand{\mx}{mumax3}
\newcommand{\mxp}{mumax\textsuperscript{+}}
\newcommand{\mxco}{MuMax-CO}

\title{\textbf{Claude-SpinDynamics: a cross-platform, dual-precision
CPU/GPU micromagnetic simulator with native \mx{} script compatibility}}

\author{%
Chun-Yeol You$^{1,*}$, Jaeyong Cho$^{1}$\\[6pt]
\normalsize
\begin{minipage}{0.88\textwidth}
\centering
$^1$Department of Physics and Chemistry, DGIST (Daegu Gyeongbuk Institute of
Science and Technology), Daegu 42988, Republic of Korea\\[3pt]
$^*$Correspondence: \href{mailto:cyyou@dgist.ac.kr}{cyyou@dgist.ac.kr}
\end{minipage}}

\date{}

\begin{document}
\maketitle

\begin{abstract}
\noindent
Quantitative spintronics increasingly depends on a handful of GPU
micromagnetic codes, all of which require NVIDIA hardware and single
precision throughout, leaving researchers without such hardware unable to
run even the standard validation problems. We report Claude-SpinDynamics
(\csd{}), a new open-source micromagnetic simulator with a cross-platform
C++20 core (Windows and Linux) and a Python interface that closes this gap:
a complete CPU build, validated by the same test suite as the GPU path, runs
every unit test and \textmu MAG standard problem with no accelerator at all,
alongside GPU builds offering both single and double precision, a choice of
two demagnetization FFT backends, and natively implemented spin--orbit,
spin-transfer, and Zhang--Li torques, Dzyaloshinskii--Moriya interaction, and
per-cell materials. \csd{} natively interprets \mx{}'s \texttt{.mx3}
scripting language, so existing community scripts run unmodified; under
matched conditions the two codes agree cell-by-cell to single-precision
round-off, and, together with \mxp{} and OOMMF, to within 2\% on the
\textmu MAG dynamic-switching standard problem, with \mxco{} agreeing to
\mx{} to float32 round-off on the same problem. Benchmarked head-to-head
against these three codes, \csd{}'s single-precision build is the fastest
solver on small and two-dimensional problems and remains competitive at
the largest grid sizes, while its double-precision and dual-FFT-backend
paths are unmatched among GPU micromagnetic codes. A GPU replica-batching
extension further
advances an entire ensemble of finite-temperature trajectories in a single
kernel launch per step, giving one to two orders of magnitude of throughput
over a per-trial loop while reproducing single-trajectory results to
numerical round-off. The complete source is openly licensed and distributed
with runnable example notebooks and documentation for independent
reproduction.
\end{abstract}

\section*{Introduction}

Micromagnetic simulation is the workhorse theory-to-device bridge of modern
spintronics. Solving the Landau--Lifshitz--Gilbert (LLG)
equation~\cite{landau1935,gilbert2004,brown1963book} on discretized geometries
is how magnetic random-access memories are scaled~\cite{dieny2017}, how
spin-transfer- and spin--orbit-torque switching channels are
engineered~\cite{ralph2008,manchon2019}, how domain-wall racetrack
devices~\cite{parkin2008} and skyrmion-based memory and logic
concepts~\cite{fert2017,muhlbauer2009,sampaio2013} are designed, and how
experiments on nanoscale textures are interpreted quantitatively when direct
imaging is impossible~\cite{fidler2000,miltat2007,abert2019,leliaert2019}.
Essentially every quantitative claim in device-scale spintronics rests on the
correctness and speed of a handful of community simulators: the CPU reference
OOMMF~\cite{oommf}, the GPU standard \mx{}~\cite{vansteenkiste2014,
leliaert2018}, the recently released extensible \mxp{}~\cite{mumaxplus} ---
a from-scratch, more modular reimplementation of the \mx{} physics engine
that adds antiferromagnetic and elastodynamic coupling but exposes its
scripting through Python rather than \texttt{.mx3} --- and
\mxco{}~\cite{mumaxco}, our own previously reported CUDA-Graph-optimized
fork of \mx{} itself (same \texttt{.mx3} scripts and physics, replaying
the per-step kernel sequence as a single graph launch for large-grid
throughput), included here as a performance baseline rather than an
independent third-party code, alongside
emerging GPU frameworks such as magnum.np~\cite{magnumnp} and
Boris~\cite{boris} and the Python finite-difference/atomistic package
Fidimag~\cite{fidimag} --- all anchored to the NIST/\textmu MAG standard
problems~\cite{mumag}.

The \mx{} family --- \mx{} itself, \mxp{}, and \mxco{} --- requires an NVIDIA
GPU to build, test, or run at all: there is no CPU fallback, so a user
without CUDA hardware cannot even execute the standard problems, and
continuous-integration testing requires GPU runners. \mx{} and its
derivatives are additionally float32-only, which
is a genuine limitation for topology-sensitive observables (skyrmion
charge, energy-landscape resolution near metastability boundaries) where
single-precision round-off is not always negligible. We built
Claude-SpinDynamics (\csd{}) to close both gaps at once: a single C++20
codebase that compiles and validates identically on Windows and Linux, with
a complete CPU build (running the full unit-test suite and every
\textmu MAG standard problem with no GPU required, for development, testing,
and teaching on any machine without CUDA hardware) alongside dual-precision GPU builds
with a choice of two demagnetization FFT backends. To lower the migration
barrier from the incumbent ecosystem, \csd{} also interprets \mx{}'s
\texttt{.mx3} scripting language natively, so existing \mx{} scripts and
workflows carry over unmodified. The GPU layer is also built against a thin,
vendor-agnostic backend seam rather than NVIDIA-specific intrinsics, so
support for AMD and Intel GPUs is a matter of implementation effort rather
than a redesign (Methods); today this is validated via a HIP backend arm
that compiles and runs through HIP-over-CUDA, with native AMD/Intel builds
available through collaboration. \csd{} was developed with the assistance of
AI-assisted coding tools under a specification-and-test discipline, following
recent demonstrations that such tools can optimize existing simulation
codes~\cite{mumaxco,you2026vrqc}; this report describes and validates the
resulting codebase.

Here we report \csd{}'s cross-platform, dual-precision, CPU-and-GPU
architecture and its native \texttt{.mx3} compatibility; validate it against
the \textmu MAG standard problems and, under matched conditions, against
\mx{} directly, including an exact cell-by-cell comparison and unmodified
\texttt{.mx3} script execution; benchmark its performance head-to-head
against \mx{}, \mxp{}, and \mxco{}, showing it is the fastest solver in the
small/2-D regime and competitive at scale while covering capabilities the
incumbent GPU codes lack; and describe a GPU replica-batching extension that
advances an entire ensemble of finite-temperature trajectories in one kernel
launch per step, with a mesh-convergence study of the switching threshold as
a worked example of the throughput it enables.

\section*{Results}

\subsection*{A cross-platform, dual-precision, CPU-and-GPU architecture}

\begin{figure}[h!]
\centering
\includegraphics[width=\textwidth]{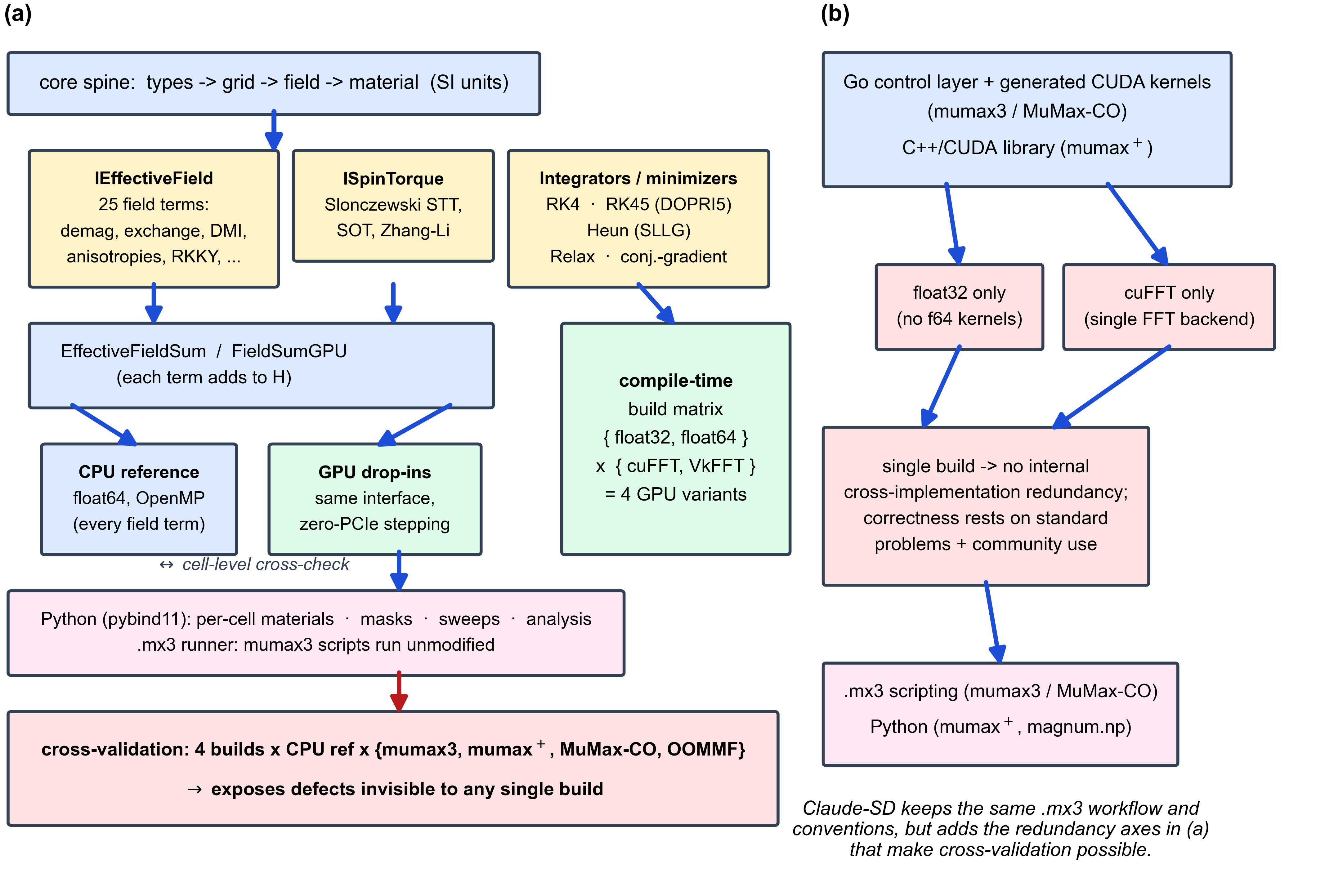}
\caption{\textbf{\csd{} code architecture versus the incumbent GPU-only
design.} (a)~\csd{}: a value-type core spine feeds narrow
\texttt{IEffectiveField}/\texttt{ISpinTorque}/integrator interfaces; every
term exists as a float64 CPU reference and, where enabled, a GPU drop-in
behind the same interface, compiling into a CPU build plus a $\{$float32,
float64$\} \times \{$cuFFT, VkFFT$\}$ GPU build matrix. (b)~The monolithic
float32/cuFFT/GPU-only \mx{} family, shown for contrast; the \texttt{.mx3}
runner keeps the incumbent scripting workflow unchanged on \csd{}.}
\label{fig:arch}
\end{figure}

\csd{} is a C++20 core (Fig.~\ref{fig:arch}) with a narrow
\texttt{IEffectiveField}/\texttt{ISpinTorque}/integrator interface: every
physics term exists as a float64 CPU reference implementation and, where a
GPU build is enabled, a CUDA drop-in behind the same interface. This
architecture, together with a Python (pybind11) interface, yields five ways
to run the same physics: a CPU library requiring no accelerator at all
(validated by the same 236-case Catch2 suite as the GPU path, so every unit
test, every \textmu MAG standard problem, and every example notebook runs on
a plain laptop), and four GPU build variants selecting float32 or float64
precision and a choice of demagnetization FFT backend: cuFFT, NVIDIA's
proprietary CUDA FFT library, or VkFFT~\cite{vkfft}, an open-source,
cross-platform (CUDA/HIP/OpenCL/Level Zero) GPU FFT implementation (125
additional GPU-tagged tests). Native effective fields and torques ---
Zeeman, uniaxial/cubic/surface anisotropy, exchange (Neumann or periodic),
open- and periodic-boundary demagnetization, bulk and interfacial DMI, RKKY
interlayer coupling, magnetoelastic anisotropy, spin--orbit torque,
Slonczewski spin-transfer torque, and Zhang--Li torque --- are first-class
fields with per-cell material control (granular media, composition
gradients, multilayer stacks), rather than user-scripted approximations.
Release GPU binaries embed kernels for compute capabilities 7.5--12.0
(Turing through Blackwell) plus forward-compatible PTX, and prebuilt Windows
packages (CPU and multi-architecture GPU) require no local build at all
(Code availability).

\subsection*{Validation against the \textmu MAG standard problems}

\begin{figure}[h!]
\centering
\includegraphics[width=0.95\textwidth]{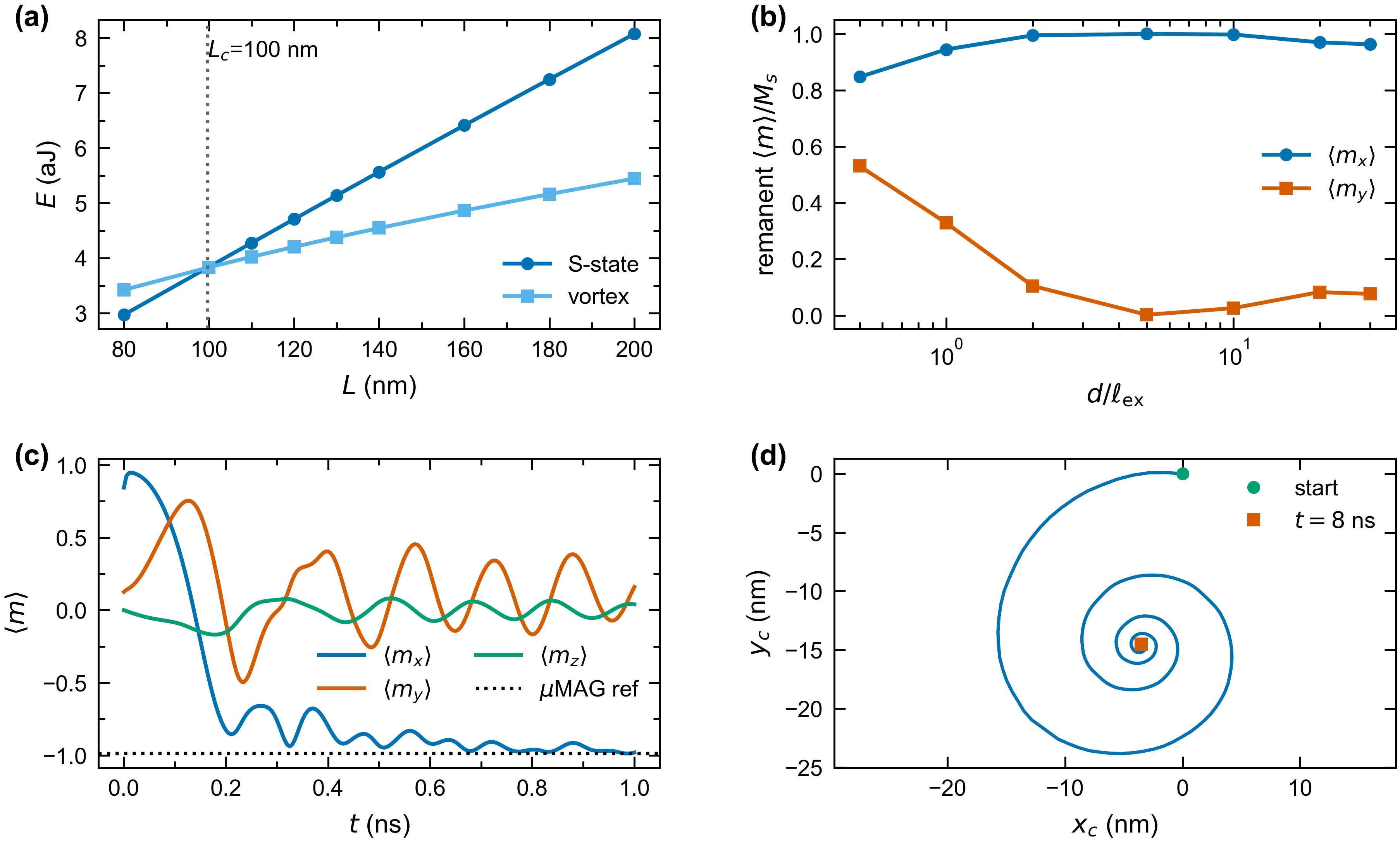}
\caption{\textbf{\textmu MAG standard-problem validation (SP\#2, SP\#4,
SP\#5) and an SP\#1-type flower/vortex crossover (\csd{}, float64).}
\csd{}'s own solution for all four panels; its agreement with \mx{},
\mxp{}, \mxco{} and the double-precision OOMMF anchor within their mutual
spread is quantified numerically in the main text (Results) and the
Supplementary Information's standard-problem tables, rather than overlaid
on these panels. \textbf{(a)} SP\#1-type: S-state/vortex energy crossing
at $L_c$ (not literally \textmu MAG SP\#1's hysteresis-loop protocol;
Supplementary Information).
\textbf{(b)} SP\#2: remanent $\langle m_x\rangle,\langle m_y\rangle$ versus
$d/\ell_\mathrm{ex}$. \textbf{(c)} SP\#4 field A: $\langle m\rangle(t)$, with
the \textmu MAG reference $\langle m_x\rangle(1\,\mathrm{ns})=-0.986$ dotted
(\csd{} value in text). \textbf{(d)} SP\#5: Zhang--Li vortex-core trajectory
over 8\,ns.}
\label{fig:umag}
\end{figure}

\csd{} reproduces the \textmu MAG standard problems SP\#2, SP\#4, and
SP\#5 in their literal geometry and protocol, agreeing
with the independent solvers within their mutual spread
(Fig.~\ref{fig:umag}, Table~\ref{tab:capability}). For the dynamic switching
problem SP\#4 (field A), $\langle m_x\rangle$ at 1\,ns is $-0.980$ (\csd{}),
$-0.980$ (\mxp{}), $-0.969$ (\mx{}) and $-0.984$ (OOMMF, CPU double), versus
the \textmu MAG reference $-0.986$ --- all within 2\%, with the
double-precision OOMMF anchor closest. On the SP\#1-type flower/vortex
crossover example below, \csd{} and \mx{} agree on
the crossover length $L_c \approx 100$--$120$\,nm (the precise
cross-code-matched sweep in the Showcase section below gives
$L_c=115.7$\,nm to both codes' agreement). For SP\#2, \csd{} and \mx{} agree
on the remanent $\langle m_x\rangle/M_s$ to $\leq 0.006$ at the four
cross-checked $d/\ell_\mathrm{ex}$ points (Supplementary Table~S2). SP\#5 reproduces the Zhang--Li
vortex-core gyration, and a separate idealized-macrospin ferromagnetic-resonance
test (no demagnetization, so no finite-size downshift; NB13, NB48) reproduces the
Kittel frequency~\cite{kittel1948} to within 0.06\% for \csd{}, \mx{}, and
\mxp{} alike; the
confined-island FMR showcase below is a distinct, geometry-resolved test
where a downshift from the Kittel value is expected.

\subsection*{Exact agreement with \mx{} under matched conditions, and native
\texttt{.mx3} script execution}

Beyond agreement within the standard problems' mutual spread, \csd{} was
built to match \mx{}'s physics conventions exactly wherever a convention is a
choice rather than a physical law, so the two codes can be compared under
genuinely identical conditions rather than merely similar ones. The
interfacial- and bulk-DMI sign, the skyrmion-polarity convention, and the
free-boundary condition $\partial\mathbf{m}/\partial n =
(D/2A)(\hat{z}\times\hat{n})\times\mathbf{m}$ at sample edges all follow
Rohart--Thiaville~\cite{rohart2013} as implemented in \mx{}, with no sign
remapping in either direction. Run under matched material parameters, cell
size, and boundary treatment, the two codes' relaxed
Dzyaloshinskii--Moriya edge-canting profiles agree cell-by-cell to
$5\times10^{-7}$ (relative), i.e.\ to single-precision round-off --- the
floor set by \mx{}'s float32 arithmetic for this comparison protocol ---
not merely ``within spread'' but numerically
the same result.

\csd{} additionally ships \texttt{micromag.mx3}, an interpreter for a
practical subset of \mx{}'s own \texttt{.mx3} scripting language
(\texttt{examples/mx3/}, 6 scripts: \texttt{sp4}, \texttt{disk},
\texttt{regions}, \texttt{relax\_demo}, \texttt{loop\_test}, and
\texttt{bext\_ac}, plus a further 12 \texttt{.mx3} scripts in
\texttt{notebooks/mx3/} used for the cross-code showcase comparisons
above), so an unmodified community \texttt{.mx3} script ---
including the standard-problem script \texttt{sp4.mx3} used throughout this
paper's benchmarking --- runs on \csd{}'s engine (CPU or GPU, auto-selected)
without translation:
\begin{quote}
\texttt{eng = mm.run\_mx3("examples/mx3/sp4.mx3", outdir="out")}
\end{quote}
or from the command line, \texttt{python -m micromag.mx3
examples/mx3/sp4.mx3}. Every benchmark comparison in this paper against \mx{}
and \mxco{} was in fact driven from identical \texttt{.mx3} scripts (Methods),
so the throughput and accuracy numbers reported below already reflect this
compatibility rather than a separately-configured native-API equivalent.
The interpreter covers the statements exercised by this paper's benchmarks
and showcase examples and by common community usage, not \mx{}'s entire
scripting API; the user guide's mumax3-to-\csd{} conversion appendix lists
supported and unsupported statements and documents the native-API
equivalent for any gap.

\subsection*{A showcase of representative examples}

Beyond the summary statistics above, we walk through seven representative
examples in full: each was solved independently with \csd{} and, where a
direct counterpart exists, with a matched \mx{} run on identical geometry,
material, and mesh. In every case the two codes agree closely; where a
residual difference exists we explain its origin rather than average over
it. All examples are distributed as runnable Python notebooks
(\texttt{notebooks/}, numbered \texttt{NB}\emph{NN}) so a reader can
reproduce every figure in this section directly (Code availability).

\begin{figure}[h!]
\centering
\includegraphics[width=\textwidth]{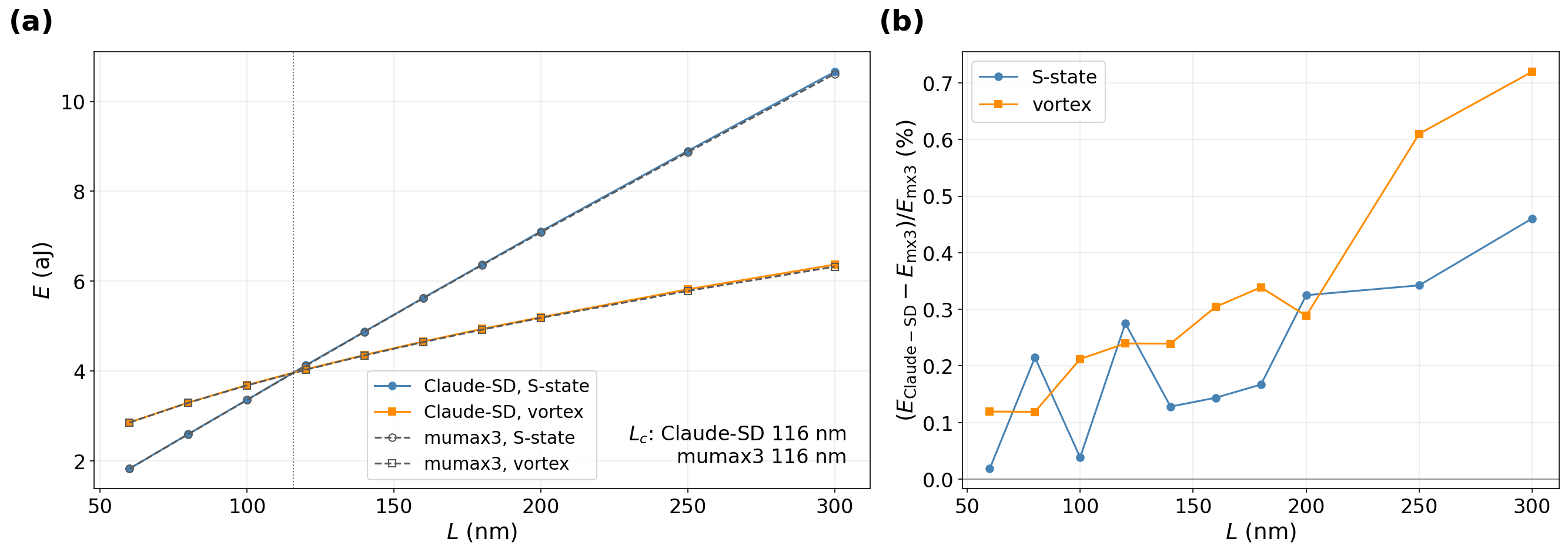}\\[4pt]
\includegraphics[width=\textwidth]{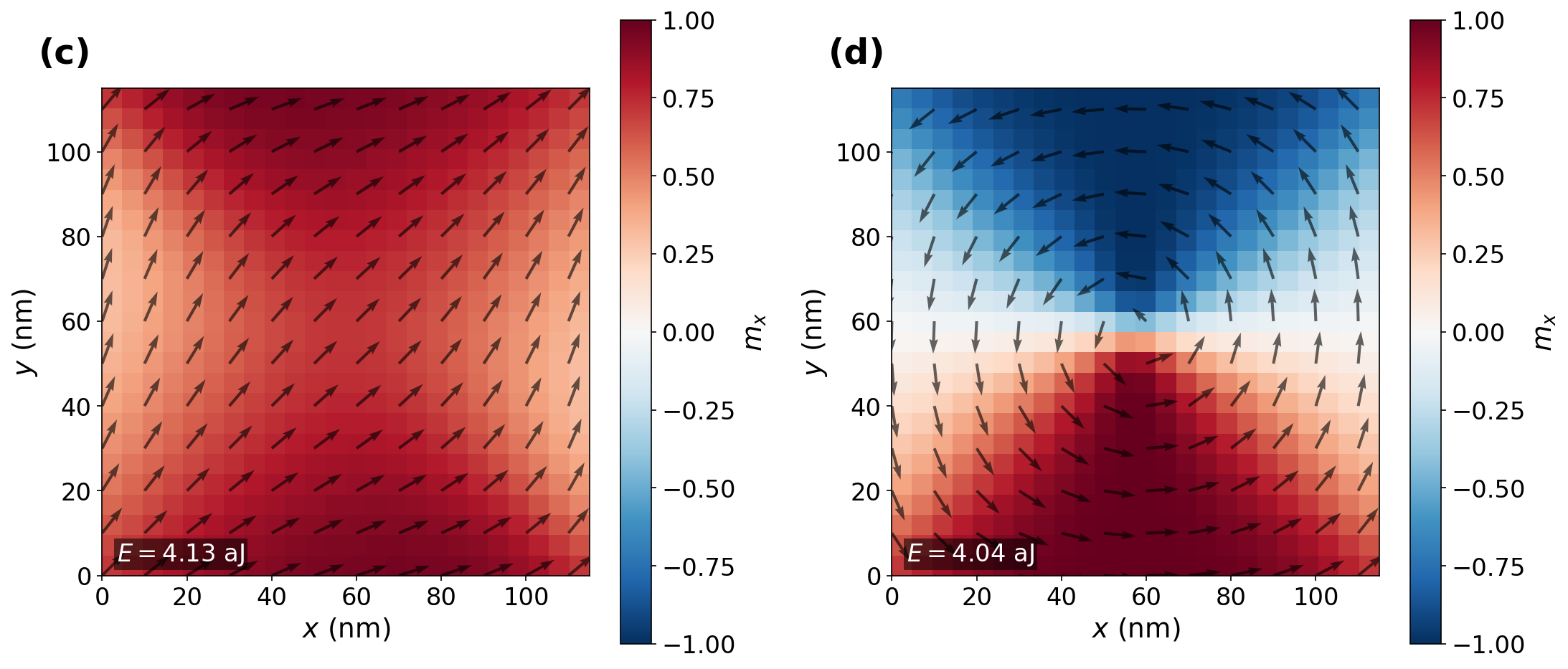}
\caption{\textbf{SP\#1-type showcase: flower/vortex crossover, \csd{} vs \mx{}.}
\textbf{(a)} Total energy of the S-state and vortex branches vs.\ element
size $L$ ($t=10$\,nm) for \csd{} and a matched \mx{} sweep (identical
$M_s$/$A_\mathrm{ex}$/$\alpha$/geometry/mesh): the crossing gives
$L_c=115.7$\,nm in both codes. \textbf{(b)} Energy difference
$(E_\mathrm{\csd{}}-E_\mathrm{mx3})/E_\mathrm{mx3}$ vs.\ $L$: $<1$\% at
every point for both branches. \textbf{(c,d)} \csd{}-only domain maps at
$L=120$\,nm: S-state ($E=4.13$\,aJ) and vortex ($E=4.04$\,aJ), the
canonical finite-size textures the crossing in (a) separates.}
\label{fig:sp1show}
\end{figure}

\paragraph{SP\#1-type flower/vortex crossover (NB02, NB46).} This example
asks for the element size at which a thin permalloy square's ground state
switches from a single-domain (``flower'' or S-state) texture to a vortex,
i.e.\ the size at which the exchange-energy cost of a vortex core is repaid
by the demagnetization-energy saved by flux closure. It uses the same
permalloy parameters and the same general element-size-dependent
validation goal as \textmu MAG SP\#1, but is not literally SP\#1's own
hysteresis-loop protocol (Supplementary Information). This is a direct
test of the exchange/demagnetization energy balance, independent of
dynamics.
Fig.~\ref{fig:sp1show}a sweeps the total energy of both branches over
element size $L$ for \csd{} and a matched \mx{} sweep (identical
$M_s$/$A_\mathrm{ex}$/$\alpha$/geometry/mesh): the crossing gives the
critical size $L_c=115.7$\,nm in \emph{both} codes, to the precision the
$20$\,nm sampling grid resolves. Fig.~\ref{fig:sp1show}b shows the
per-point energy difference is $<1$\% throughout, with no systematic trend
in $L$ --- the residual is consistent with ordinary relaxation-tolerance
noise, not a physics discrepancy. Fig.~\ref{fig:sp1show}c,d show the two
competing textures directly (S-state and vortex, $L=120$\,nm, \csd{}
only): the canonical finite-size domain structures the problem is testing
for.

\begin{figure}[h!]
\centering
\includegraphics[width=\textwidth]{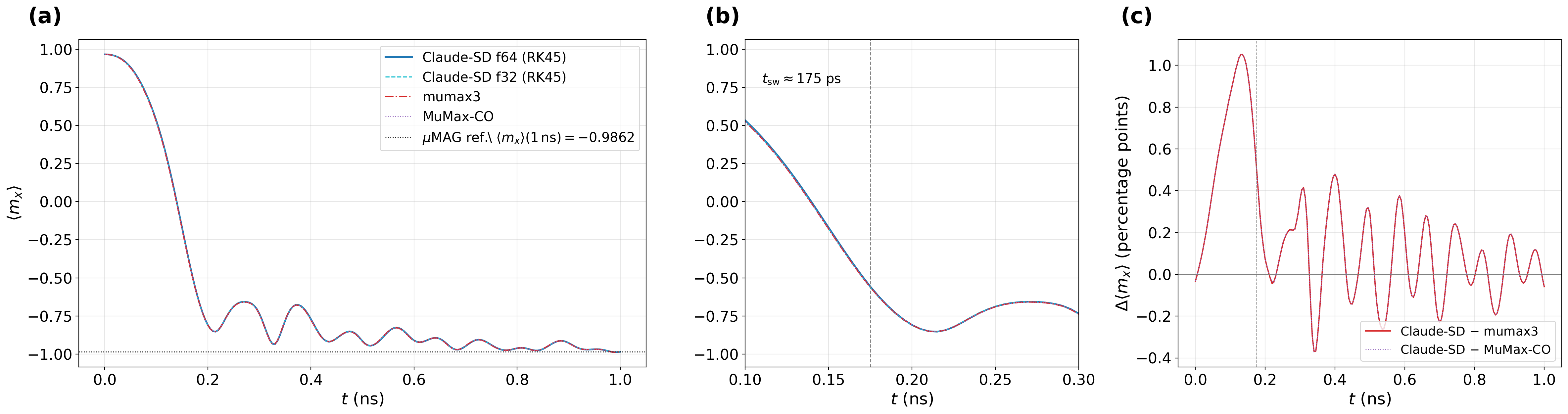}
\caption{\textbf{SP\#4 showcase: field-driven switching dynamics, \csd{}
vs \mx{} vs MuMax-CO.} \textbf{(a)} $\langle m_x\rangle(t)$ over the full
1\,ns reversal of the $500\times125\times3$\,nm permalloy strip,
\csd{} float64/float32 (adaptive RK45) vs.\ \mx{} vs.\ MuMax-CO, with the
\textmu MAG consensus $\langle m_x\rangle(1\,\mathrm{ns})=-0.9862$
(dotted). \textbf{(b)} Zoom on the switching region; all solvers cross
$\langle m_x\rangle=0$ at $t_\mathrm{sw}\approx175$\,ps. \textbf{(c)}
Pointwise difference $\langle m_x\rangle_\mathrm{\csd{}}-\langle
m_x\rangle_\mathrm{mumax3}$ (and vs.\ MuMax-CO): $<1.1$ percentage points
throughout, largest at the switching transient.}
\label{fig:sp4show}
\end{figure}

\paragraph{SP\#4 --- field-driven switching dynamics (NB01, NB41).} Unlike
the SP\#1-type example above, SP\#4 is a genuinely \emph{dynamic} benchmark: a reversing
field is applied to a thin permalloy strip and the full precessional
switching trajectory --- not just the final state --- must be integrated
correctly. Fig.~\ref{fig:sp4show}a overlays \csd{} (float64/float32,
adaptive RK45) against \mx{} and MuMax-CO across the full 1\,ns trajectory;
Fig.~\ref{fig:sp4show}b zooms on the switching event itself
($t_\mathrm{sw}\approx175$\,ps in all solvers). Fig.~\ref{fig:sp4show}c
makes the agreement quantitative: the pointwise difference stays below 1.1
percentage points for the entire trajectory, with the largest deviation ---
still a small fraction of the trajectory's full $[-1,1]$ range --- occurring
transiently right at the switching event, where the
trajectories are most sensitive to last-bit timing differences. This
complements the single-endpoint numbers already quoted above
($\langle m_x\rangle$ at 1\,ns) with the full time-resolved picture.

\begin{figure}[h!]
\centering
\includegraphics[width=\textwidth]{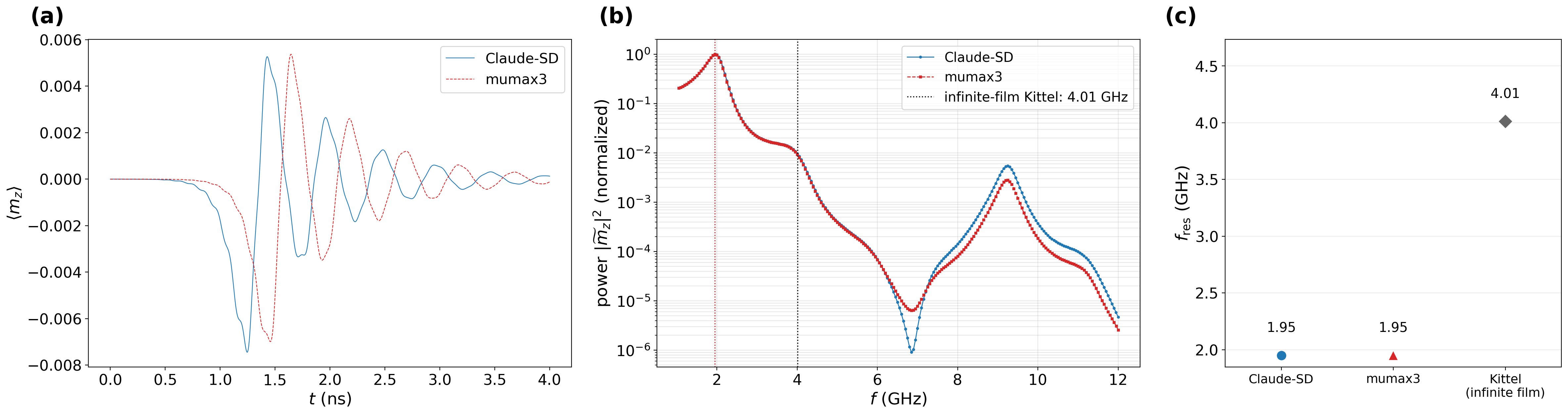}
\caption{\textbf{FMR showcase: broadband ferromagnetic resonance of an
$80\times80\times4$\,nm permalloy island, \csd{} vs a matched \mx{} run.}
\textbf{(a)} $\langle m_z\rangle(t)$ ring-down ($B_\mathrm{bias}=20$\,mT
in-plane) --- \csd{} driven by a continuous sinc-field pulse, \mx{} by a
brief broadband kick on the identical geometry/material; the excitation
waveforms differ, so the time-domain envelopes differ, but both probe the
same normal mode. \textbf{(b)} Power spectrum (20\,ns record,
$\Delta f=0.05$\,GHz): the island's quasi-uniform mode lies below the
infinite-film Kittel value~\cite{kittel1948} because of finite-size
in-plane demagnetization, reproduced identically by both codes.
\textbf{(c)} Extracted resonance frequencies: \csd{} and \mx{} coincide
to within one $0.05$\,GHz spectral bin.}
\label{fig:fmrshow}
\end{figure}

\paragraph{Ferromagnetic resonance of a confined island (NB13, NB48).}
Broadband FMR is a standard dynamic-response benchmark distinct from the
\textmu MAG suite: a sinc- (\csd{}) or impulse-excited (\mx{}) $80\times
80\times4$\,nm permalloy island is left to ring down freely, and the
resonance frequency is extracted from the power spectrum
(Fig.~\ref{fig:fmrshow}a,b). Because the two codes use different broadband
excitation waveforms, their time-domain ring-down envelopes differ
(Fig.~\ref{fig:fmrshow}a) --- this is expected and does not indicate a
physics disagreement, since both waveforms are designed only to excite the
same set of normal modes broadband, not to reproduce each other's
transient. The power spectra confirm this: extracted resonance
frequencies coincide to within one $0.05$\,GHz spectral bin (the two
peaks fall in the same FFT bin of a 20\,ns record; no sub-bin
interpolation was applied, Fig.~\ref{fig:fmrshow}c), and both sit below the
infinite-film Kittel value~\cite{kittel1948} by the same
confined-geometry demagnetization downshift.

\begin{figure}[h!]
\centering
\includegraphics[width=\textwidth]{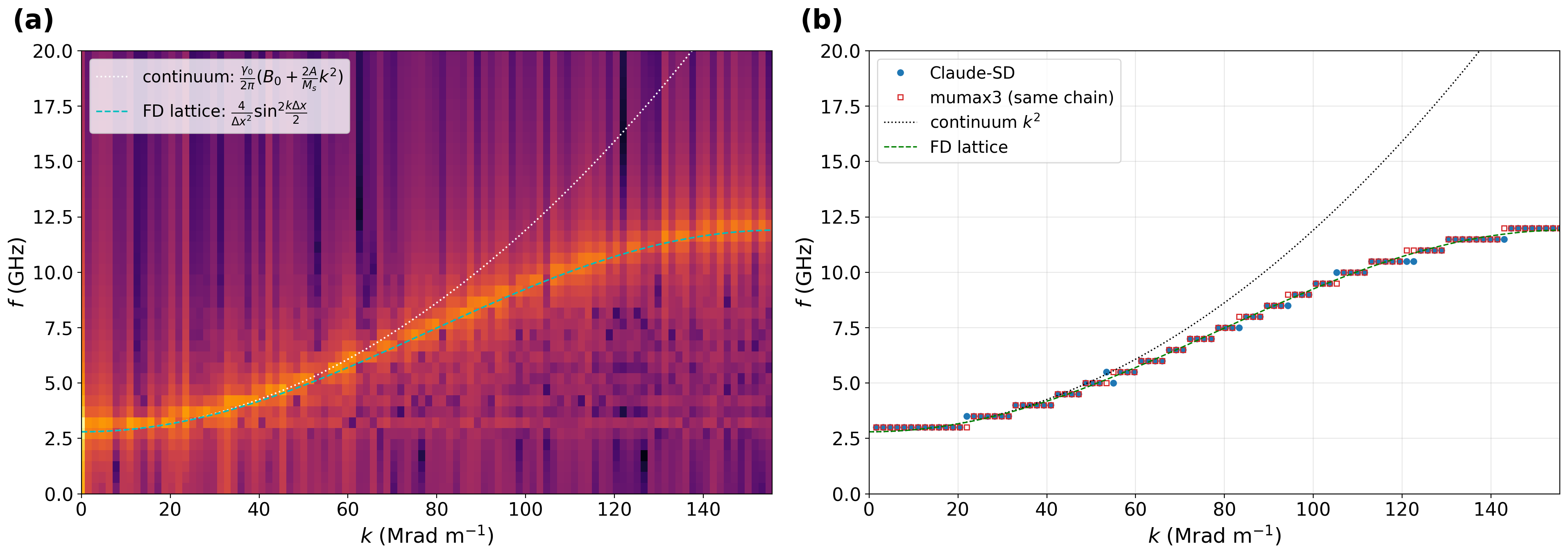}
\caption{\textbf{Spin-wave dispersion showcase: \csd{} vs \mx{} vs
lattice theory.} 1-D permalloy chain ($200\times1\times1$,
$\Delta x=20$\,nm, $B_0=0.1$\,T, demag off; $\Delta x$ exceeds the
permalloy exchange length $\ell_\mathrm{ex}\approx5.7$\,nm by design, since
this test targets the discrete lattice Laplacian's own dispersion, not
continuum-limit micromagnetic accuracy, and demagnetization is switched
off so both codes solve the identical exchange-only dispersion relation
exactly). \textbf{(a)} \csd{} $S(k,f)$
map with the continuum parabola $f=\frac{\gamma_0}{2\pi}(B_0+\frac{2A}
{M_s}k^2)$ (white) and the exact finite-difference lattice dispersion
$f=\frac{\gamma_0}{2\pi}\!\left[B_0+\frac{2A}{M_s}\frac{4}{\Delta
x^2}\sin^2\!\frac{k\Delta x}{2}\right]$ (cyan). \textbf{(b)} Extracted
peak positions --- \csd{} and \mx{} (identical chain, point-antenna
excitation) coincide with each other and with the lattice curve at every
$k$; both depart from the continuum parabola above
$k\approx60$\,Mrad\,m$^{-1}$, the expected finite-difference-lattice
signature, not a code discrepancy.}
\label{fig:dispshow}
\end{figure}

\paragraph{Exchange spin-wave dispersion (NB15).} A 1-D permalloy chain
excited by a point antenna maps out the magnon dispersion relation
$f(k)$ (Fig.~\ref{fig:dispshow}a). \csd{} and \mx{} extracted peak
positions coincide with each other \emph{and} with the exact
finite-difference lattice dispersion at every $k$
(Fig.~\ref{fig:dispshow}b); both depart from the continuum parabola
$f\propto k^2$ above $k\approx60$\,Mrad\,m$^{-1}$, saturating toward the
Brillouin-zone edge. This departure is the expected signature of solving
the lattice Laplacian on a finite-difference mesh (which replaces $k^2$ by
$\frac{4}{\Delta x^2}\sin^2(k\Delta x/2)$), not a numerical error in
either code --- the continuum formula is only valid for $k\Delta x\ll1$.

\begin{figure}[h!]
\centering
\includegraphics[width=\textwidth]{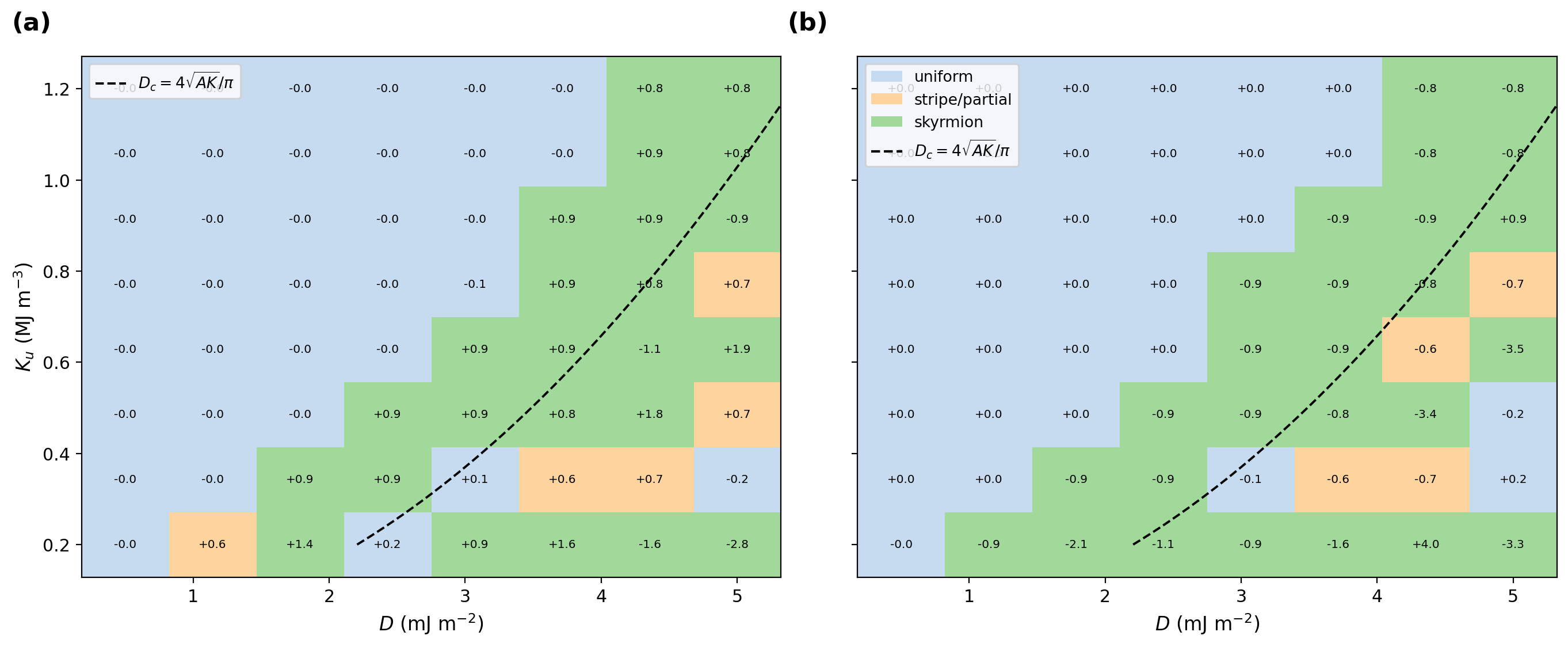}
\caption{\textbf{Skyrmion $D$--$K$ phase diagram under matched dynamics,
\csd{} vs \mx{}: 92\% cell-level agreement.} Pt/Co film
($192\times192\times3$\,nm), identical N\'eel-skyrmion seed, both codes
integrating the \emph{full} Landau--Lifshitz--Gilbert equation
(precession and damping) at the same Gilbert damping $\alpha=0.05$ for
the same physical time (30\,ns) --- \csd{} with \texttt{RK45IntegratorGPU}
(adaptive DOPRI5) (a), \mx{} with \texttt{alpha=0.05; Run(30ns)} (b).
Cells show the relaxed topological charge $Q$ (sign as computed by each
code), classified uniform/stripe/skyrmion by $|Q|$, with the analytic
boundary $D_c=4\sqrt{AK}/\pi$ overlaid. The seeded skyrmion survives in a
band that tracks $D_c$ in both codes, and the cell-by-cell phase
classification matches between \csd{} and \mx{} at 92\% of grid points;
the residual 8\% sits only in the low-$K_u$/high-$D$ corner, where the
ground state is a multi-$Q$ stripe/labyrinth whose exact winding is
chaotic and seed-sensitive in any solver. This 92\% is sign-insensitive
(uniform/stripe/skyrmion classification only): the relaxed core polarity
itself can differ between the two codes at a given grid point, an
initial-condition/basin effect rather than a sign-convention mismatch
(main text).}
\label{fig:phaseshow}
\end{figure}

\paragraph{Skyrmion $D$--$K$ phase diagram under matched dynamics
(NB21).} A seeded N\'eel skyrmion in a Pt/Co film
($192\times192\times3$\,nm) is relaxed across a grid of DMI strength $D$
and perpendicular anisotropy $K_u$, and the relaxed topological charge
$Q$ classifies each cell as uniform, stripe/partial, or skyrmion
(Fig.~\ref{fig:phaseshow}), tracing out the phase boundary against the
analytic estimate $D_c=4\sqrt{AK}/\pi$. When both codes integrate the
\emph{full} Landau--Lifshitz--Gilbert equation (precession and damping) at
the same Gilbert damping $\alpha=0.05$ for the same physical time
(30\,ns) --- \csd{} with \texttt{RK45IntegratorGPU} (adaptive DOPRI5),
\mx{} with \texttt{alpha=0.05; Run(30ns)} --- the seeded skyrmion survives
in a band that tracks $D_c$ in both codes, and the cell-by-cell phase
classification (uniform/stripe/skyrmion, from $|Q|$) matches between \csd{}
and \mx{} at 92\% of grid points; the residual 8\% sits only in the
low-$K_u$/high-$D$ corner, where the ground state is a multi-$Q$
stripe/labyrinth whose exact winding is chaotic and seed-sensitive in any
solver. The sign of $Q$ itself is not compared cell-by-cell here: the two
codes' relaxed skyrmions can settle into opposite core polarity from the
same nominally-identical seed --- an initial-condition/basin choice, not a
convention mismatch (Methods documents where sign conventions are matched
exactly, e.g.\ the DMI edge-canting comparison in the Supplementary
Information) --- so the 92\% figure is a sign-insensitive phase-diagram
agreement, not a claim that every cell's signed $Q$ coincides. This is an
algorithm-dependent metastability effect: pairing \csd{}'s damped-LLG
relaxer or energy minimizer against \mx{}'s
conjugate-gradient minimizer on this same $D$--$K$ sweep gives only
53--56\% cell-level agreement, because the two relaxation operators can
settle into different metastable (multi-)skyrmion basins near the phase
boundary; matching the operator class --- both codes integrating the same
damped dynamics --- raises the agreement to 92\%. The lesson generalizes
beyond this one example: near a metastability boundary, the relaxation
operator is part of the physics being compared, and cross-code agreement
should be assessed protocol-by-protocol rather than assumed from a single
run.

\paragraph{Replica-batched examples (NB30--32; see Fig.~\ref{fig:batchstt},
Fig.~\ref{fig:batchmesh} below).} The remaining two examples ---
thermally-assisted STT switching statistics and the MTJ mesh-convergence
study --- exercise the replica-batched engine described below and have no
direct \mx{} counterpart, since batched-ensemble execution is a \csd{}
capability rather than a shared benchmark; they are validated instead
against analytic theory (finite-window N\'eel--Brown) and against
themselves across mesh refinement. We present them together with the
replica-batching architecture in the next subsection rather than
duplicating the figures here.

Two further cross-code examples --- current-driven domain-wall motion and
Walker breakdown, and N\'eel-skyrmion DMI edge canting --- are presented
in the Supplementary Information with the same level of detail; both show
the same pattern of close \csd{}/\mx{} agreement once conventions and
solver classes are matched.

\subsection*{Capability and performance: a four-solver comparison}

\begin{figure}[h!]
\centering
\includegraphics[width=0.9\textwidth]{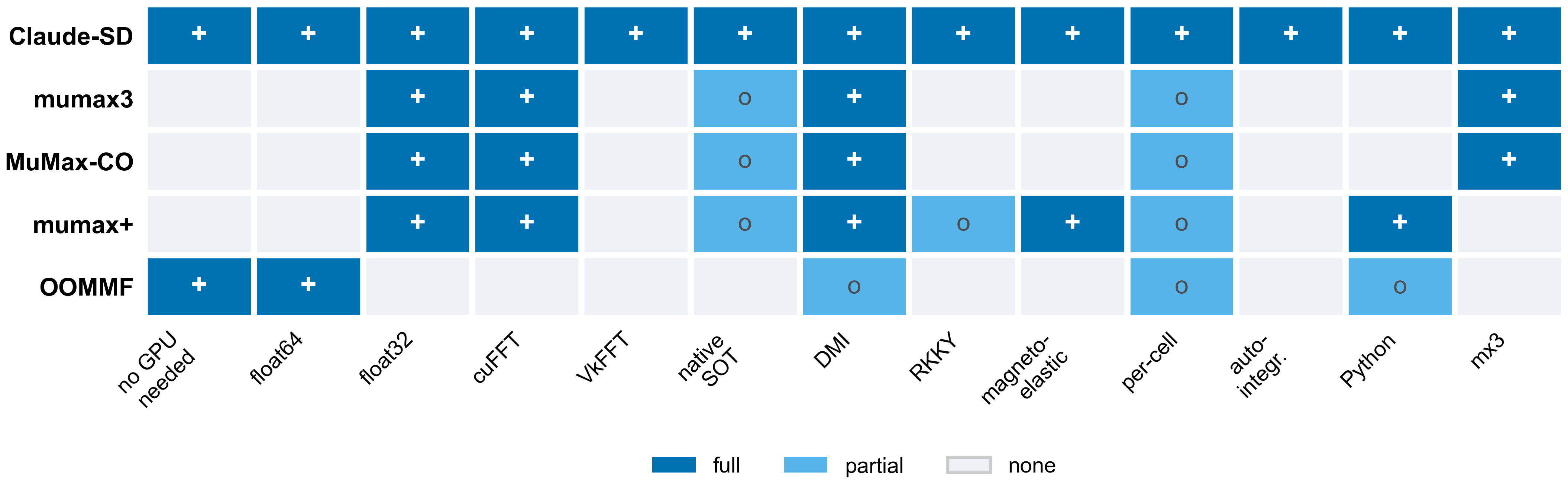}
\caption{\textbf{Solver capability matrix.} Only \csd{} covers every axis:
a CPU build requiring no GPU, double precision, a second FFT backend
(VkFFT), native SOT, RKKY, magnetoelastic coupling, per-cell materials, and
automatic integrator selection.}
\label{fig:capability}
\end{figure}

We benchmarked \csd{} against \mx{}~\cite{vansteenkiste2014}, \mxp{}
(mumaxplus 1.2.1)~\cite{mumaxplus}, and \mxco{}~\cite{mumaxco} --- our own
previously reported CUDA-Graph-optimized \mx{} fork, included as a
performance baseline rather than an independent code --- on an NVIDIA RTX
5060 Ti (Blackwell, CUDA 13.2). The solvers differ first in capability (Table~\ref{tab:capability},
Fig.~\ref{fig:capability}): the entire \mx{} family is float32-only and
GPU-only, whereas \csd{} offers float32 \emph{and} float64 on both CPU and
GPU; \csd{} alone carries two demag FFT backends (cuFFT and VkFFT), native
spin--orbit torque~\cite{manchon2019}, per-cell materials, and an automatic
integrator selector. For a spintronics workload this capability gap is not
cosmetic: chiral textures governed by the Dzyaloshinskii--Moriya
interaction~\cite{dzyaloshinskii1958,moriya1960,rohart2013},
spin--orbit-torque switching~\cite{manchon2019}, and Zhang--Li domain-wall
motion~\cite{zhang2004,thiaville2012} are exactly the problems where
topology-sensitive observables benefit from double precision, and where
per-cell material control (granular media, composition gradients, multilayer
stacks) is required. We therefore compare \emph{speed} at float32 (the only
common precision) and anchor \emph{accuracy} to \csd{}'s float64 build.

\begin{figure}[h!]
\centering
\includegraphics[width=0.85\textwidth]{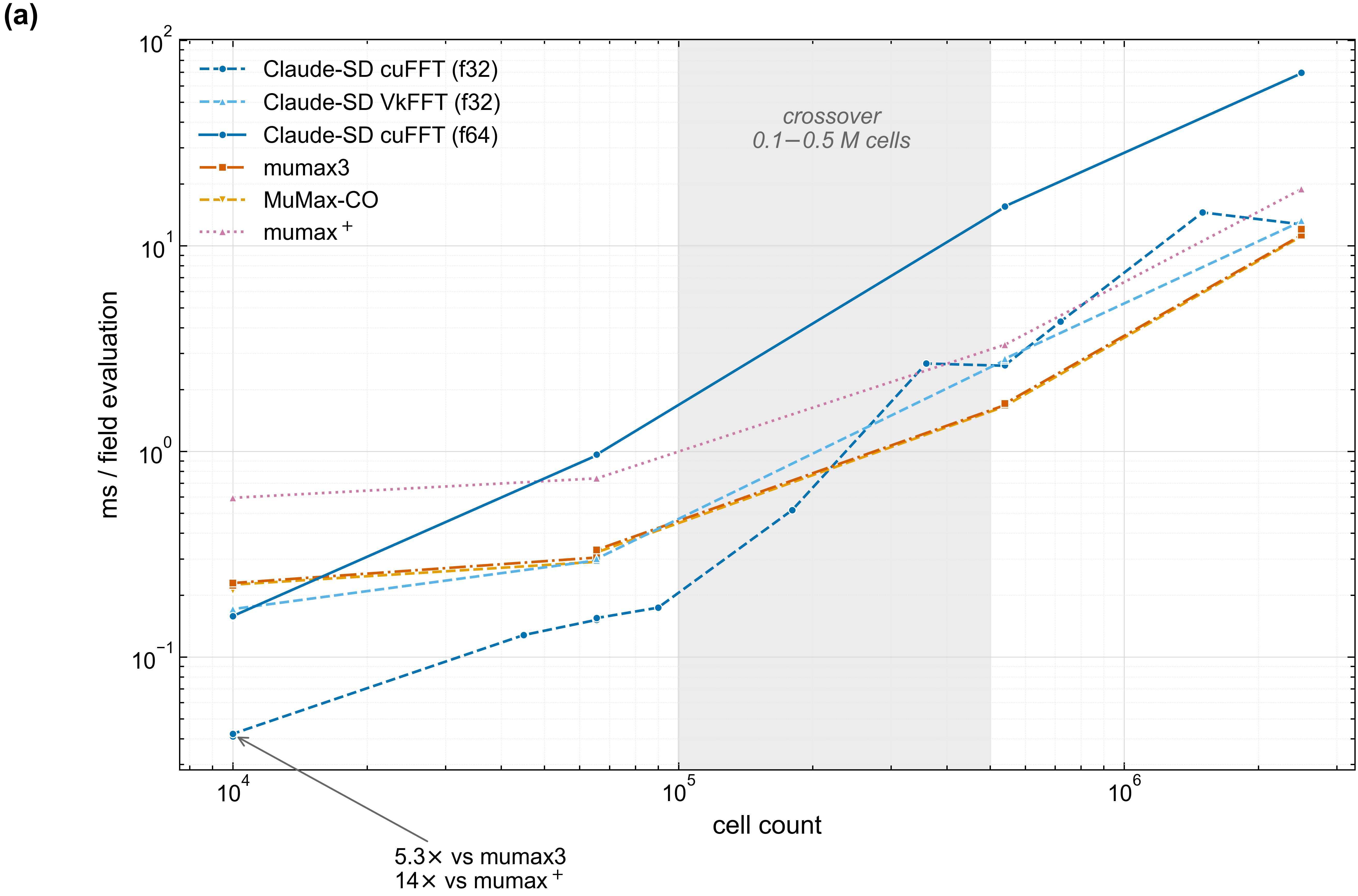}\\[4pt]
\includegraphics[width=0.85\textwidth]{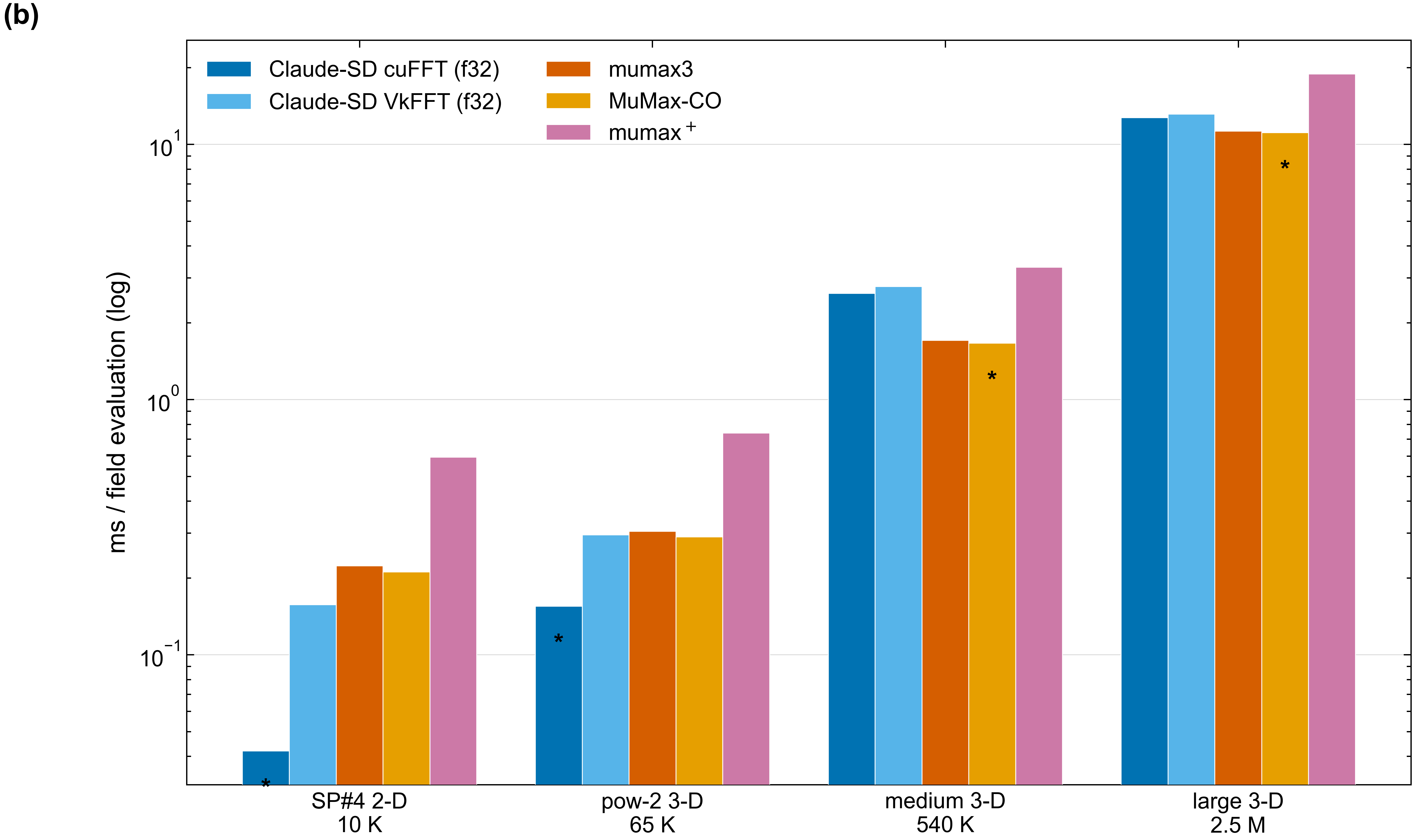}
\caption{\textbf{Four-solver throughput.} (a)~ms per field evaluation versus
cell count: \csd{} float32 leads below the $\approx$0.1--0.5\,M-cell
crossover; the \mx{} family leads above it. (b)~Per-scenario bars for the
four benchmark grids (Table~\ref{tab:throughput}); the asterisk above a bar
marks the fastest solver for that scenario.}
\label{fig:throughput}
\end{figure}

Because the solvers use different-order integrators (RK4 = 4,
Dormand--Prince~\cite{dormand1980} = 6, Heun = 2 field evaluations per step),
the fair throughput metric is milliseconds per field evaluation (Methods).
Table~\ref{tab:throughput} and Fig.~\ref{fig:throughput} give the result.
Four findings follow.

\textbf{(1) A crossover at $\approx$0.1--0.5\,M cells.} Below it,
kernel-launch overhead dominates and \csd{}'s cuFFT float32 build --- which
captures the per-step kernel sequence as a CUDA Graph~\cite{cudagraphs} and
replays it --- is the fastest solver: 5.3$\times$ faster than \mx{} and
14$\times$ faster than \mxp{} at the SP\#4 grid, and 2.0$\times$/4.8$\times$
at 65\,K cells. Above the crossover the comparison becomes cuFFT-bound and the
\mx{} family leads, with \csd{} within 1.57$\times$ at 540\,K cells and
1.14$\times$ at 2.5\,M cells.

\textbf{(2) \mxco{} is the fastest large-grid solver} and matches or beats
stock \mx{} everywhere: its CUDA-Graph replay removes launch overhead on
small grids and reaches parity on large ones (the graph path auto-disables
beyond $\approx$0.8\,M cells). On the full SP\#4 1-ns \emph{adaptive} run
(integration phase, setup-subtracted), \mxco{} completes 3.2$\times$ faster
than stock \mx{} (5.2\,s $\to$ 1.6\,s), with final states agreeing to
float32 round-off.

\textbf{(3) \mxp{} is the slowest GPU solver on raw throughput} at every
size; its value lies in extensibility (antiferromagnets, elastodynamics), not
speed.

\textbf{(4) The dual-precision advantage is quantitative.} Within \csd{},
float32 is 4--6$\times$ faster than float64 at large 3-D sizes on the
Blackwell Tensor-Core FFT (e.g.\ $61.8 \rightarrow 10.4$\,ms/step at 540\,K
cells; $276 \rightarrow 51$\,ms/step at 2.5\,M), and VkFFT overtakes cuFFT for
large float64 transforms while losing on small grids --- so the two backends
are complementary. No competing GPU code exposes a second FFT backend, and
none of the codes benchmarked here exposes a precision choice; among GPU
micromagnetic codes more broadly, the PyTorch-based magnum.np~\cite{magnumnp}
can switch tensor dtype, but without precision-specialized kernels or an
alternative FFT backend.

The finite-temperature picture mirrors $T=0$: with matched
Stratonovich--Heun stochastic-LLG integrators~\cite{brown1963,garcia1998}
(the only fair SLLG comparison, since \mxco{}'s graph path is disabled at
$T>0$ and reduces to \mx{}), \csd{} is 2.1$\times$ faster than \mx{} at the
SP\#4 grid and $\approx$parity at 0.2\,M cells.

The comparison is also platform-robust. \csd{} builds and validates on both
Windows/MSVC and native Linux/GCC; on a same-host Linux comparison (NVIDIA
L4, Ada) the small-grid advantage over \mx{} persists (2.4$\times$ at the
SP\#4 grid, parity at 65\,K cells) with the same crossover shape, and it is
unchanged when the workload is extended from demag+exchange to a realistic
chiral-film field set (interfacial DMI + perpendicular anisotropy). The
small-grid win is therefore neither a Windows artifact nor an artifact of a
minimal field set.

\textbf{Where each code wins.} No single solver dominates on speed
(Fig.~\ref{fig:throughput}), and on capability only \csd{} covers every axis
(Fig.~\ref{fig:capability}). \csd{} is preferable for small/2-D dynamics, for
any study that needs double-precision reference accuracy or a
topology-sensitive observable, for development or teaching on a machine
without a GPU, and for native custom-torque (SOT/DMI/Zhang--Li) or
per-cell-material physics scripted from Python; \mxco{}/\mx{} are preferable
for large float32 production; \mxp{} for antiferromagnetic or magnetoelastic
problems. That \csd{} is competitive with --- and on small problems faster
than --- hand-tuned incumbent codes, while adding capabilities they lack, is
the central performance result.

\subsection*{A replica-batched engine for finite-temperature ensembles}

\begin{figure}[h!]
\centering
\includegraphics[width=0.95\textwidth]{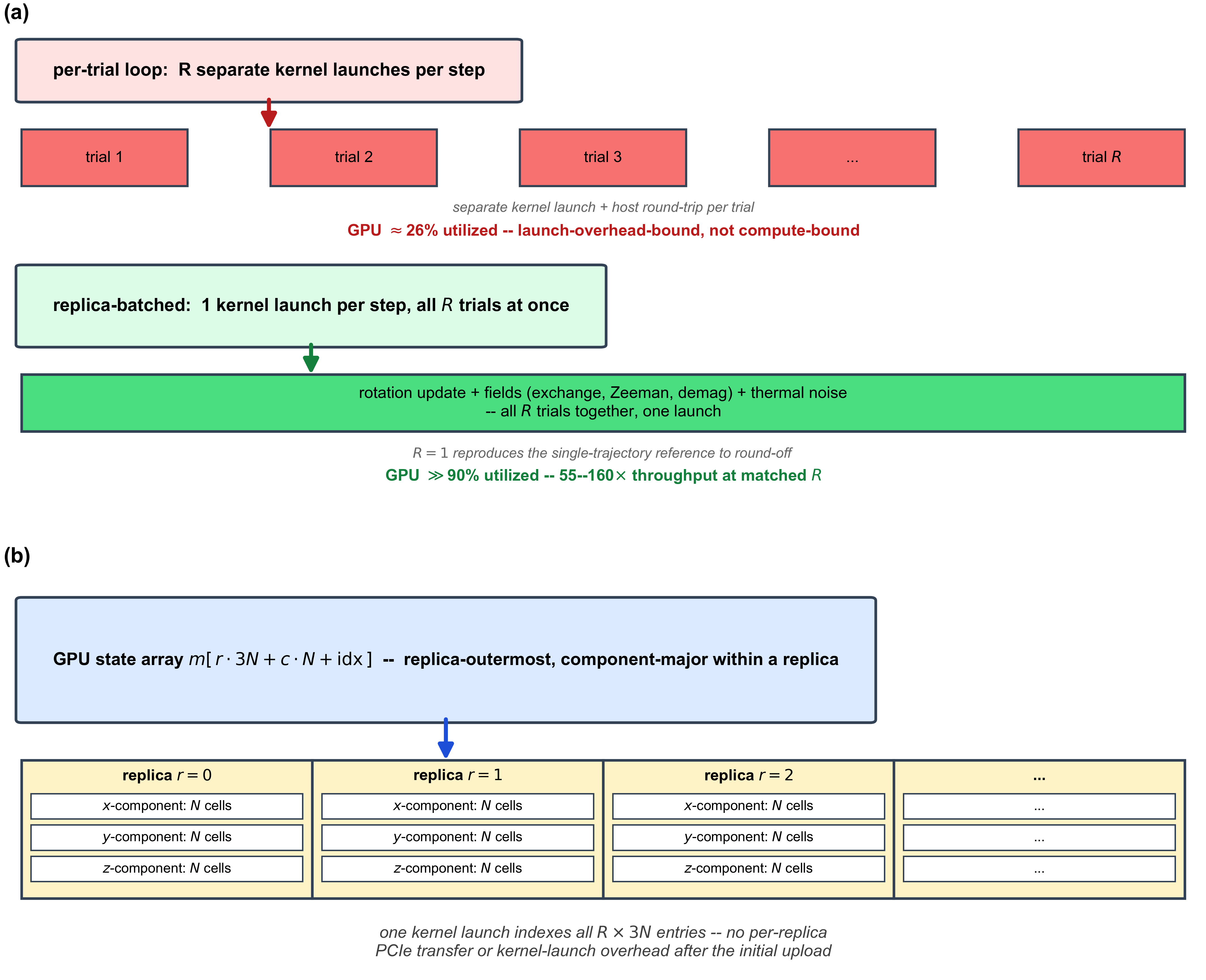}
\caption{\textbf{The replica-batching execution model.} (a)~A per-trial loop
issues $R$ separate kernel launches per step, one host round-trip each,
leaving the GPU $\approx$26\% utilized; replica-batching instead appends a
replica dimension so rotation update, all fields, and thermal noise for all
$R$ trials advance together in a single kernel launch per step, at
$\gg$90\% utilization and 55--160$\times$ the per-trial-loop throughput at
matched $R$. $R{=}1$ reproduces the single-trajectory reference integrator
to round-off, so batching changes only performance, not physics.
(b)~The corresponding GPU state layout: replica-outermost,
component-major within each replica ($m[r\cdot 3N + c\cdot N +
\mathrm{idx}]$), so one kernel launch indexes all $R\times 3N$ entries with
no per-replica PCIe transfer after the initial upload.}
\label{fig:replica}
\end{figure}

Stochastic switching-probability and bit-error-rate studies --- the workload
behind thermally-assisted spin-transfer-torque (STT) switching and MRAM
reliability estimation --- require hundreds to thousands of independent
finite-temperature trajectories per operating point. Run one trial at a
time, each launch leaves the GPU $\approx$26\% utilized (Fig.~\ref{fig:replica}a): the workload is
launch-overhead-, not compute-, bound. \csd{} addresses this with a
GPU-resident \emph{replica-batching} layer (Fig.~\ref{fig:replica}a): a
leading replica dimension $R$ is prepended to the magnetization state
(replica-outermost, component-major layout, Fig.~\ref{fig:replica}b), and
every kernel --- rotation-based integration, exchange/uniaxial/Zeeman
fields, demagnetization, and thermal noise --- advances all $R$ trials in
one launch per step, each trial carrying its own current density $J$ and
temperature $T$.

Integration uses a norm-exact Depondt--Mertens rotation
scheme~\cite{depondt2009} (Rodrigues update, $|\mathbf{m}|=1$ to round-off),
verified so that $R=1$ reproduces the single-trajectory reference
integrator to numerical round-off (max.\ difference $5.9\times10^{-15}$ for
local fields, $1.3\times10^{-10}$ for the demagnetization path, consistent
with double-precision FFT round-off). Thermal noise uses a device-side
counter-based Philox stream keyed by (seed, replica, step), giving
statistically independent, bit-reproducible realizations per trial; the
finite-temperature thermal-field variance
$\sigma = \sqrt{2\alpha k_BT/(\mu_0^2 M_sV\gamma_0\Delta t)}$ was validated
against the parameter-free Langevin equilibrium law $\langle m_z\rangle =
\mathcal{L}(\xi)$, $\xi = \mu_0 M_sVH/k_BT$: $\langle m_z\rangle = 0.675$
versus $\mathcal{L}(3)=0.672$. An optional retire/refill mechanism freezes a
trial the step it crosses a stop condition (e.g.\ switched) and reactivates
its slot with a fresh noise stream, giving variable-length trials without
starving GPU occupancy.

Table~\ref{tab:batching} quantifies the result on a single-cell macrospin
(local fields, STT, thermal noise), an $8\times8\times1$
exchange-and-Zeeman grid, and a $16\times16\times1$ grid with the full
demagnetization path added: this controlled microbenchmark, which isolates pure
kernel throughput at matched total work (same physics, same replica count,
looped launches vs.\ one batched launch), gives 55--160$\times$, scaling to
$3.7\times10^7$ replica$\times$step/s at $R=1024$. On a full
thermally-assisted Slonczewski-STT switching sweep (a Pt/Co-like macrospin,
10{,}000 replicas $\times$ 20{,}000 steps covering 25 current-density points
with 400 trials each), the end-to-end wall time --- including the
per-trial Python driver and GPU-object setup that the microbenchmark above
factors out --- drops from $\approx$2.5\,h for a naive per-trial loop to
$\approx$0.6\,s batched, a larger ratio than Table~\ref{tab:batching}'s
kernel-only figure because the per-trial baseline here also pays
non-kernel overhead that the batched path amortizes away entirely; the
55--160$\times$ range is the conservative, apples-to-apples number we
report as the batching speedup proper. The
batched switching probability (Fig.~\ref{fig:batchstt}) reproduces the
sigmoidal shape of the finite-window N\'eel--Brown~\cite{brown1963} law
with Sun's antidamping barrier reduction~\cite{sun2000},
$P_\mathrm{sw}=1-\exp[-tf_0e^{-\Delta(1-J/J_{c0})^2}]$ (clamped at
$J\geq J_{c0}$, with $f_0$ and $\Delta$ fixed independently from the
macrospin parameters, no free fit), but the zero-fit-parameter theory
curve crosses $P_\mathrm{sw}=0.5$ at $J/J_{c0}\approx0.85$ while the
batched simulation crosses it at $J/J_{c0}\approx0.97$: Sun's linearized
barrier-reduction form is an asymptotic approximation that is known to
lose accuracy close to the threshold current itself, where the underlying
mean-field treatment of the antidamping torque is least exact, so a
threshold offset of this size against the simulated (not merely the
analytic) instability point is an expected model limitation rather than a
solver discrepancy; the independently measured $T=0$ instability current
$J_{c0}$ that both curves are normalized by is matched exactly by
construction.

The speedup makes a discretization-convergence study of the switching
threshold practical (Fig.~\ref{fig:batchmesh}): the same
$80\times80\times1.5$\,nm CoFeB free layer, meshed at
$8\times8$/$16\times16$/$32\times32$/$64\times64$ cells
(10/5/2.5/1.25\,nm), gives a switching threshold that converges only once
the cell size drops below the exchange length
$\ell_\mathrm{ex}=\sqrt{2A/\mu_0M_s^2}=4.89$\,nm: the 16-, 32-, and 64-cell
meshes agree on $J_{c0}=0.60\times10^{12}$\,A\,m$^{-2}$ and the
fully-switched threshold $J(P_\mathrm{sw}{=}1)=0.63\times10^{12}$\,A\,m$^{-2}$,
while the coarser $8\times8$ mesh ($10$\,nm $>\ell_\mathrm{ex}$)
under-resolves the reversal and is biased $\approx$25\% high
($J_{c0}=0.75$ vs.\ $0.60\times10^{12}$\,A\,m$^{-2}$). Spatially
resolving the free layer on the converged $16\times16$ mesh (exchange +
demagnetization + Slonczewski STT + thermal noise, versus a coherent
whole-layer macrospin at matched $T=0$ threshold) also shows a broader,
higher-overdrive transition consistent with nucleation-mediated reversal
rather than coherent rotation --- physics invisible to the macrospin
approximation.

\subsection*{Open availability: examples, notebooks, and documentation}

The complete source, unit tests, and benchmark harness are distributed under
the GPLv3 license (Code availability). Beyond the library itself, the
repository ships 51 runnable Python example notebooks spanning every
\textmu MAG standard problem, ferromagnetic resonance, spin-wave dispersion,
domain-wall motion, skyrmion nucleation and phase diagrams, Walker breakdown,
spin--orbit-torque switching, and the finite-temperature replica-batching
studies reported above; 9 further API examples (including a
ParaView-visualization gallery); and 18 \texttt{.mx3} scripts runnable both
on \mx{} and on \csd{}'s own engine (6 general-purpose examples in
\texttt{examples/mx3/} plus 12 used to drive this paper's cross-code
showcase figures, \texttt{notebooks/mx3/}). A 1200-line user guide documents the full
Python API, GPU per-cell field interface, integrator-selection guidance,
and a mumax3-to-\csd{} conversion appendix, so a reader can reproduce every
figure in this paper, or start a new study, without reading the C++ source.

\section*{Discussion}

\csd{} demonstrates that a cross-platform, dual-precision micromagnetic
simulator can match the incumbent GPU standard exactly under matched
conditions, run its scripts unmodified, and still be faster in the small/2-D
regime while adding capabilities the incumbent lacks.

Relative to \mx{} --- the de facto GPU standard of the spintronics
community~\cite{vansteenkiste2014,leliaert2018} --- \csd{} is differentiated
on five axes. (i)~\emph{A CPU build with no GPU requirement}: every code we
benchmark head-to-head here (\mx{}, \mxp{}, \mxco{}) requires an NVIDIA GPU
to build or run at all; \csd{}'s CPU path, validated by the same native-kernel
test suite as its GPU path, makes development, teaching, and small-scale
production possible on any machine.
(ii)~\emph{Double precision}: the \mx{} family is float32-only; \csd{}'s
float64 path enables bit-reproducible topological-charge studies, long-time
integrations, and energy-landscape resolution that single precision cannot
guarantee, at a measured and acceptable cost (float32 remains available and
4--6$\times$ faster when precision permits). (iii)~\emph{Redundant demag
backends}: cuFFT and VkFFT are complementary in size regime and provide an
internal cross-check of the single most error-prone component of any
micromagnetic code. (iv)~\emph{Native spintronics physics}: spin--orbit
torque, Zhang--Li torque, bulk and interfacial DMI, RKKY interlayer coupling,
and magnetoelastic anisotropy are first-class fields rather than
user-scripted approximations, with per-cell material control for granular and
graded media. (v)~\emph{A Python-native API with \mx{} compatibility}: \csd{}
embeds in the scientific-Python ecosystem (NumPy interchange, and a
multiprocessing-safe \texttt{parameter\_sweep} helper that can distribute
independent runs across multiple GPUs via per-worker device selection)
while also executing \mx{} \texttt{.mx3} scripts
unmodified and matching \mx{} to $5\times10^{-7}$ under matched conditions,
lowering the migration barrier in both directions. \mx{} retains genuine
advantages --- a decade of community hardening, a large published-results
corpus, and the best large-grid float32 FFT throughput --- and our data
quantify rather than dismiss them. Among the codes not benchmarked here, the
PyTorch-based magnum.np~\cite{magnumnp} is the closest in capability breadth
(including dtype selection and a comparable native-physics set) and is
uniquely differentiable via autograd for inverse design; its own benchmarks
place it within a factor of two of \mx{} at large sizes, while the
launch-overhead-dominated small-grid regime --- where \csd{}'s advantage is
largest --- is a known cost of eager tensor-framework execution. Unlike the
\mx{} family, magnum.np is PyTorch-based and so can, like \csd{}, run its
CPU tensor backend without a GPU; \csd{}'s CPU path differs in being
validated by the identical native-kernel test suite used for the GPU path,
rather than inheriting CPU support from a general-purpose tensor framework.
The two designs are complementary: differentiable simulation versus
maximum-throughput, cross-validated native kernels.

The replica-batched extension adds a capability none of the benchmarked
solvers offers: an entire finite-temperature ensemble advanced in one launch,
gated by an $R=1$ round-off-identical regression against the single-trajectory
integrator on every new kernel. It also surfaced a numerical-accuracy lesson
worth stating plainly: because the exchange field scales as $1/dx^2$,
refining the mesh without shrinking the time step accordingly does not crash
--- the norm-exact rotation integrator stays stable --- it silently
suppresses switching and returns a smooth but wrong curve. Finally, because
the GPU kernels use no vendor-specific intrinsics, the same backend seam that
isolates CUDA from the physics code also isolates it from the vendor: a HIP
backend arm compiles and executes through HIP-over-CUDA today, and native
AMD/Intel builds are available through collaboration, extending the reach of
the codebase beyond the single-vendor GPU ecosystem that constrains \mx{},
\mxp{}, and \mxco{} alike.

The limitations are bounded honestly. \csd{}'s damped-LLG relaxer converges
more slowly than adaptive minimizers for stiff Dzyaloshinskii--Moriya
textures; topological-charge studies near phase boundaries should use the
energy minimizer or a larger step budget. The raw-throughput advantage is
confined to the small/2-D regime; large float32 production remains the \mx{}
family's domain. The replica-batched demagnetization kernel uses a
full-complex (not symmetry-compressed) tensor contraction --- simpler and
already correct at the single-cell-to-few-thousand-cell scale the batching
targets, but not yet the production kernel's memory-optimized form at large
per-replica grids --- and no other benchmarked solver here offers a
comparable batching feature, so this extension is reported as a capability
advance rather than a cross-solver comparison. The \texttt{.mx3} interpreter
covers a practical subset of \mx{}'s scripting language (Appendix A of the
user guide lists supported and unsupported statements), not its entirety.

\section*{Methods}

\subsection*{Physics and discretization}
\csd{} integrates
\begin{equation}
\frac{d\mathbf{m}}{dt} =
 -\gamma'\mu_0\,(\mathbf{m}\times\mathbf{H})
 -\gamma'\alpha\mu_0\,\mathbf{m}\times(\mathbf{m}\times\mathbf{H}),
\qquad \gamma' = \frac{\gamma_0}{1+\alpha^2},
\end{equation}
on a structured finite-difference grid with $x$-fastest indexing.
Effective-field contributions (each adds to $\mathbf{H}$): Zeeman,
uniaxial/cubic/surface anisotropy, six-point Laplacian exchange (Neumann or
periodic boundary conditions), open-boundary demagnetization by FFT
convolution with the Newell tensor~\cite{newell1993} zero-padded to $2N$,
periodic-boundary demagnetization by an image-sum kernel, bulk and
interfacial DMI, RKKY interlayer exchange, and magnetoelastic anisotropy. The
interfacial-DMI sign and skyrmion-polarity conventions follow
Rohart--Thiaville~\cite{rohart2013} as implemented in \mx{}, including the
free boundary condition $\partial\mathbf{m}/\partial n =
(D/2A)(\hat{z}\times\hat{n})\times\mathbf{m}$ at sample edges (grid and
geometry-mask boundaries). One convention difference from \mx{} is
documented for reproducibility: the Zhang--Li drift velocity is
$u = JP\mu_B/(eM_s)$ (without Thiaville's $1/(1+\xi^2)$
prefactor~\cite{thiaville2012} that \mx{} includes). Because Walker
breakdown occurs at a fixed dimensionless drift velocity $u_c$ set by
$\alpha$ and $\xi$ alone, the two conventions map the same physical
current $J$ to drift velocities $u$ differing by $1+\xi^2$, so both the
sub-Walker drift velocity at a given $J$ \emph{and} the Walker breakdown
current $J_W$ itself differ by $1+\xi^2$ between the two conventions
unless matched; an option matches \mx{} exactly when required (used
throughout the Zhang--Li showcase and Supplementary comparisons). Spin torques (Slonczewski
STT~\cite{slonczewski1996,berger1996}, spin--orbit torque~\cite{manchon2019},
Zhang--Li~\cite{zhang2004}) are added per integrator stage. Finite-temperature
dynamics use the stochastic LLG with a Stratonovich Heun scheme and cuRAND
thermal fields~\cite{brown1963,garcia1998}, or the replica-batched
Depondt--Mertens scheme (below).

\subsection*{GPU implementation}
Effective fields and integrators run entirely on device with no per-step host
transfer; a field compositor accumulates contributions either on per-field
streams (synchronized between fields) or on a single shared stream
(serialized by stream order). Demagnetization uses cuFFT or, optionally,
VkFFT~\cite{vkfft}; the CPU reference path uses FFTW~\cite{fftw}, the
long-standing open-source C FFT library on which most CPU micromagnetic
codes (including OOMMF) rely. Fixed-step
integrators capture the per-step kernel sequence as a CUDA
Graph~\cite{cudagraphs} and replay it. Single and double precision are
selected at build time (four preset GPU builds: cuFFT/VkFFT $\times$
f32/f64, plus the CPU build). Release binaries embed kernels for compute
capabilities 7.5--12.0 (Turing through Blackwell) plus forward-compatible
PTX, and the runtime verifies kernel--device compatibility at import.

\paragraph{Portability beyond NVIDIA.}
The GPU layer is written against a thin compile-time backend seam (runtime,
kernel launch, FFT, and RNG), and the kernels use no vendor-specific
intrinsics (no warp-shuffle/ballot or texture memory), so a port to other
vendors is mechanical rather than structural. A HIP backend arm is provided
for AMD (ROCm); its runtime, launch, and RNG paths compile and execute
through HIP-over-CUDA (\texttt{hipcc}, \texttt{HIP\_PLATFORM=nvidia}),
validating the seam on a second backend without AMD hardware. An Intel path
via SYCL (AdaptiveCpp/DPC++), which additionally runs on CPU, is planned.
Native AMD and Intel builds are available through collaboration on request.

\subsection*{Integrators}
RK4 (fixed step), RK45 Dormand--Prince (adaptive, first-same-as-last)
\cite{dormand1980}, Heun (stochastic LLG), a norm-exact Depondt--Mertens
rotation scheme~\cite{depondt2009} (Rodrigues update; optionally adaptive),
damping-only \texttt{RelaxGPU}, and a Barzilai--Borwein/FIRE
\texttt{MinimizeGPU}. \texttt{recommend\_integrator()} selects among them
from damping, temperature, goal, and an analytic phase-error estimate, and
logs the matched \mx{} \texttt{SetSolver} index so cross-solver runs use
equivalent schemes.

\subsection*{Replica batching}
The finite-temperature ensemble engine (Results) adds a leading replica
dimension $R$ to the Depondt--Mertens integrator: state buffers are laid out
replica-outermost and component-major within a replica, local-field, thermal,
and demagnetization kernels are re-indexed over $R\times N$ threads, and
demagnetization batches the forward/inverse FFT over $3R$ transforms with the
Newell kernel tensor~\cite{newell1993} precomputed once and shared across
replicas (identical geometry). Current density $J$ and temperature $T$ are
per-replica arrays, giving independent sweep points in a single launch.
Thermal noise uses a device-side counter-based Philox generator keyed by
(seed, replica-and-cell index, step), reused identically between the
predictor and corrector sub-steps (Stratonovich). Every kernel is verified by
an $R=1$ regression against the single-trajectory
\texttt{DepondtMertensGPU} reference trajectory before being accepted.

\subsection*{Benchmark protocol}
Throughput was measured by two-run subtraction,
$[t(N_2)-t(N_1)]/(N_2-N_1)$, to cancel one-time set-up, with size-tiered step
counts so the timed difference is several seconds of compute; five measured
repeats are reported as median with interquartile range, on an idle RTX 5060
Ti. A \texttt{cudaDeviceSynchronize} precedes every clock stop. Solvers use
different-order integrators, so the fair metric is ms per field evaluation
(RK4 = 4, Dormand--Prince = 6, Heun = 2 evaluations per step). \mx{} and
\mxco{} were driven from identical \texttt{.mx3} scripts (Results); \mxp{}
through its Python \texttt{TimeSolver}; OOMMF via \texttt{boxsi}. The Linux
cross-solver comparison used the same protocol on an NVIDIA L4 (Ada, Ubuntu,
CUDA 12.4) against the official \mx{} 3.12 Linux binary; the \mx{}
3.11.1$\to$3.12 version difference was measured on the primary GPU and is
$\leq 8$\% at the smallest grid and $\leq 1$\% elsewhere. All measurements
share one record schema in \texttt{benchmarks/results/all\_solvers.json};
\texttt{make\_report.py} regenerates every table and figure from it.

\subsection*{Software availability and reproducibility}
The use of AI-assisted coding tools in developing \csd{} is disclosed in
accordance with the Nature Portfolio policy on large language models: the
authors specified the architecture, physics conventions, and acceptance
tests, reviewed every change, and take full responsibility for the code and
this manuscript; no LLM is listed as an author. Build presets, environment,
exact commands, and per-scenario grids are in the Supplementary Information;
the full benchmark is scripted end-to-end and regenerates all tables and
figures from a single command.

\section*{Data availability}
All data supporting the findings of this study --- benchmark records
(\url{benchmarks/results/all_solvers.json}), \textmu MAG validation
outputs, and the scripts that regenerate every table and figure --- are
available in the code repository at
\url{https://github.com/mirryou-maker/Claude-SpinDynamics} and will be
archived on Zenodo with a DOI upon publication.

\section*{Code availability}
The complete source code of \csd{}, including the CPU reference
implementations, CUDA kernels, unit tests, and the four-solver benchmark
harness, is available under the GPLv3 license at
\url{https://github.com/mirryou-maker/Claude-SpinDynamics} (archived on
Zenodo; citation metadata in \texttt{CITATION.cff}). Prebuilt Windows
binaries (CPU and multi-architecture GPU packages, requiring no local build)
are provided via GitHub Releases; a 1200-line user guide
(\texttt{docs/USER\_GUIDE.md}) documents the full API. Beyond the figures
shown in this paper and its Supplementary Information, the same GitHub
repository ships substantially more worked examples than are reproduced
here: 51 runnable Python notebooks under \texttt{notebooks/} (catalogued
in Supplementary Table~S8 and in \texttt{notebooks/GALLERY.md}, covering
every \textmu MAG standard problem, FMR, spin-wave dispersion,
domain-wall and skyrmion dynamics, Walker breakdown, spin--orbit-torque
switching, and the replica-batched finite-temperature studies), 9 further
API examples, and 18 \texttt{.mx3} scripts runnable unmodified on either
\csd{} or \mx{} (6 general-purpose plus 12 driving this paper's own
showcase figures); every one of them regenerates its own figure(s) against
the current build and is a starting point for further exploration.


\section*{Acknowledgements}
This work was supported by the National Research Foundation of Korea (NRF)
(No.~RS-2026-25502724) and the Strategic Research Program under the DGIST
R\&D Program (26-SR-01) of the Ministry of Science, ICT, and Future Planning.

\section*{Author contributions}
C.-Y.Y. conceived the study, directed development, designed the validation
and benchmark protocols, analyzed the data, and wrote the manuscript. J.C.
contributed to validation and reviewed the manuscript.

\section*{Competing interests}
The authors declare no competing interests.

\section*{Ethics declarations}
\textbf{Ethics approval and consent to participate.} Not applicable: this
study did not involve human participants, human data or tissue, or animal
subjects.\\
\textbf{Consent for publication.} Not applicable.

\section*{Materials availability}
This study did not generate physical materials. All computational
materials --- source code, build configurations, and benchmark/validation
data --- are covered under Code availability and Data availability above.

\section*{Supplementary information}
Supplementary Information is included as an ancillary file with this
arXiv submission, and is also available with the published version of
this article, including extended benchmark methodology, full
standard-problem protocol details, the complete precision/race-condition
study, the AI-agent development protocol, additional cross-code
validation figures (Walker breakdown and DMI edge canting), and the full
example-notebook catalog.

\clearpage
\section*{Tables}

\begin{table}[h!]
\centering
\caption{\textbf{Solver capability.} \csd{} alone offers a CPU build
requiring no GPU, dual precision, a second FFT backend, natively implemented
spin--orbit torque and per-cell materials, and an automatic integrator
selector. Scripting: Py = Python API, C++ = native library, mx3 = \mx{}
\texttt{.mx3} interpreter. $^\dagger$\mx{} has no first-class RKKY field;
inter-layer exchange coupling is instead scripted via its
\texttt{ext\_InterExchange}/\texttt{ext\_ScaleExchange} region-rescaling
mechanism.}
\label{tab:capability}
\footnotesize
\begin{tabularx}{\textwidth}{@{} l c c c X c @{}}
\toprule
 & Runs w/o GPU & Precision & Demag FFT & Native SOT/DMI/RKKY/ME & Scripting \\
\midrule
\textbf{\csd{}} & \textbf{yes (CPU)} & \textbf{f32 + f64} & \textbf{cuFFT \& VkFFT} &
  \textbf{all native} + per-cell + auto-select & Py + C++ + mx3 \\
\mx{}   & no & f32 & cuFFT & DMI; STT via Slonczewski, no native SOT; no RKKY$^\dagger$ & .mx3 \\
\mxco{} & no & f32 & cuFFT & = \mx{} (+ CUDA-Graph path) & .mx3 \\
\mxp{}  & no & f32 & cuFFT & DMI, \textbf{AFM, elastodynamics} & Py \\
\bottomrule
\end{tabularx}
\end{table}

\begin{table}[h!]
\centering
\caption{\textbf{Throughput, ms per field evaluation} (float32, $T=0$; lower
is better; median of five repeats). CS = \csd{}. Bold marks the fastest solver per row.}
\label{tab:throughput}
\footnotesize
\begin{tabular}{@{}llcccccc@{}}
\toprule
Scenario & cells & dim & CS cuFFT\_f32 & CS VkFFT\_f32 & \mx{} & \mxco{} & \mxp{} \\
\midrule
SP\#4 2-D   & 10\,K  & 2D & \textbf{0.042} & 0.157 & 0.223 & 0.211 & 0.595 \\
pow-2 3-D   & 65\,K  & 3D & \textbf{0.155} & 0.295 & 0.305 & 0.290 & 0.739 \\
medium 3-D  & 540\,K & 3D & 2.610 & 2.777 & 1.708 & \textbf{1.663} & 3.302 \\
large 3-D   & 2.5\,M & 3D & 12.72 & 13.16 & 11.28 & \textbf{11.13} & 18.87 \\
\bottomrule
\end{tabular}
\end{table}

\begin{table}[h!]
\centering
\caption{\textbf{Replica-batching throughput} (\csd{} only; no comparable
feature exists in \mx{}, \mxp{}, or \mxco{}). Speedup is one batched launch of
$R$ replicas versus $R$ sequential per-trial launches, at matched total work;
$R=1$ reproduces the single-trajectory reference trajectory to round-off in
every case (Methods).}
\label{tab:batching}
\footnotesize
\begin{tabularx}{\textwidth}{@{} X c c c c @{}}
\toprule
Physics & Grid & $R$ & Speedup vs.\ loop & Throughput at $R{=}1024$ \\
\midrule
Macrospin (local fields + STT + thermal) & $1\times1\times1$ & 256 & 158.7$\times$ & $3.7\times10^7$ replica$\cdot$step/s \\
Multi-cell (exchange + uniaxial + Zeeman) & $8\times8\times1$ & 64 & 59.8$\times$ & $1.4\times10^8$ cell$\cdot$step/s ($R{=}256$) \\
Multi-cell + demagnetization & $16\times16\times1$ & 64 & 55.4$\times$ & $8.9\times10^7$ cell$\cdot$step/s ($R{=}256$) \\
\bottomrule
\end{tabularx}
\end{table}

\clearpage
\section*{Figure legends}

\begin{figure}[h!]
\centering
\includegraphics[width=0.72\textwidth]{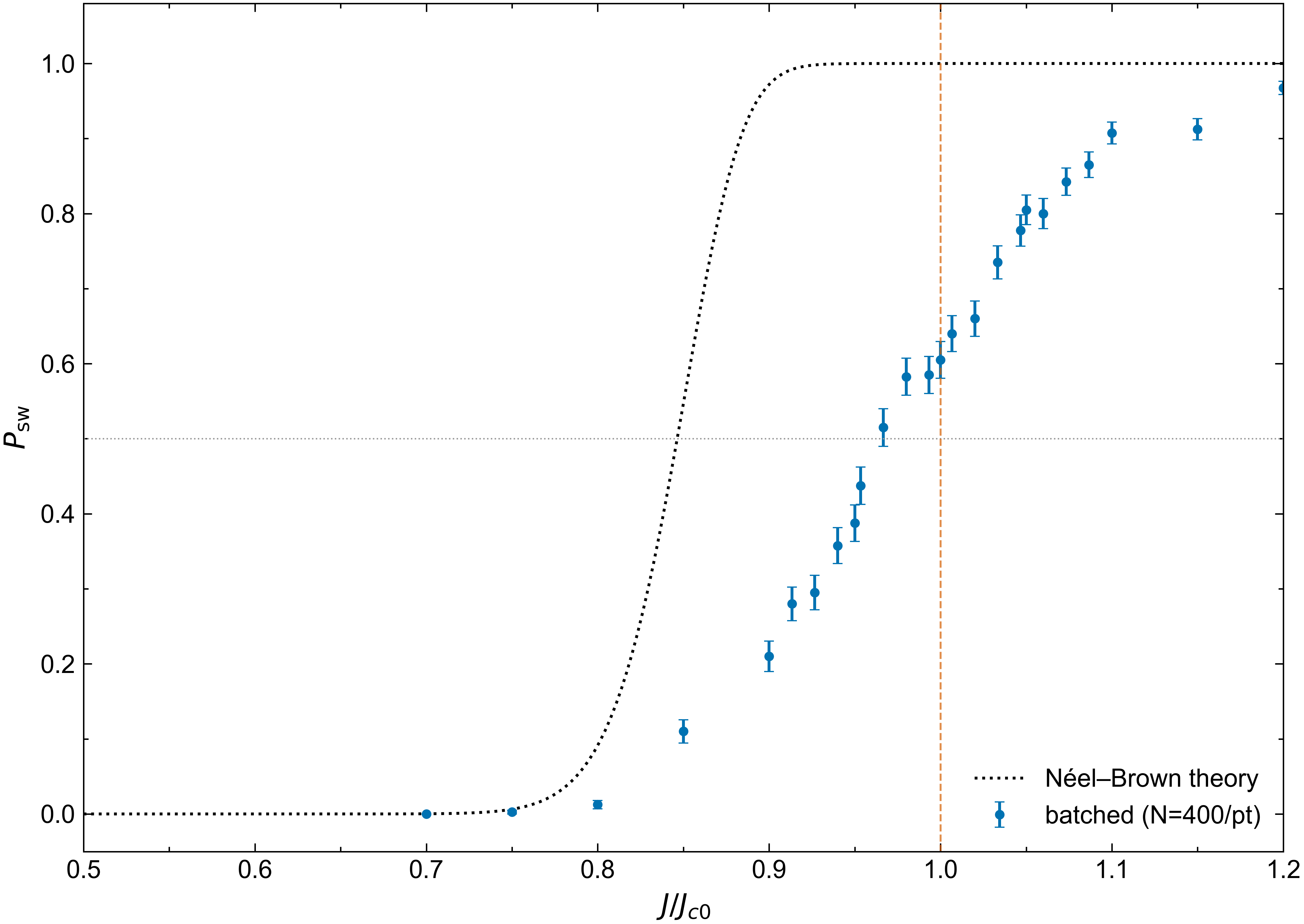}
\caption{\textbf{Replica-batched thermally-assisted STT switching.} Switching
probability of a Pt/Co macrospin (10\,nm cell, thermal stability
$\Delta_{300\mathrm{K}}=KV/k_BT=121$, instability threshold
$J_{c0}=2.60\times10^{12}$\,A\,m$^{-2}$) versus normalized current
$J/J_{c0}$: 10{,}000 GPU replicas (400 trials/point, 25 points), advanced by one
kernel launch per step,
complete the full sweep in $\approx$0.6\,s versus $\approx$2.5\,h for a
naive per-trial loop (end-to-end wall time, including per-trial driver
overhead absorbed by batching; Table~\ref{tab:batching}'s matched-work kernel-only
comparison gives the more conservative 55--160$\times$). The batched result reproduces the sigmoidal shape of the finite-window
N\'eel--Brown~\cite{brown1963} law with Sun's barrier
reduction~\cite{sun2000} (dashed, zero free parameters) but crosses
$P_\mathrm{sw}=0.5$ at a $\approx$14\% higher $J/J_{c0}$ than the
analytic curve, an expected limitation of Sun's linearized threshold
approximation (main text); the $T=0$ threshold $J_{c0}$ is measured
independently from a small-tilt seed (main text).}
\label{fig:batchstt}
\end{figure}

\begin{figure}[h!]
\centering
\includegraphics[width=\textwidth]{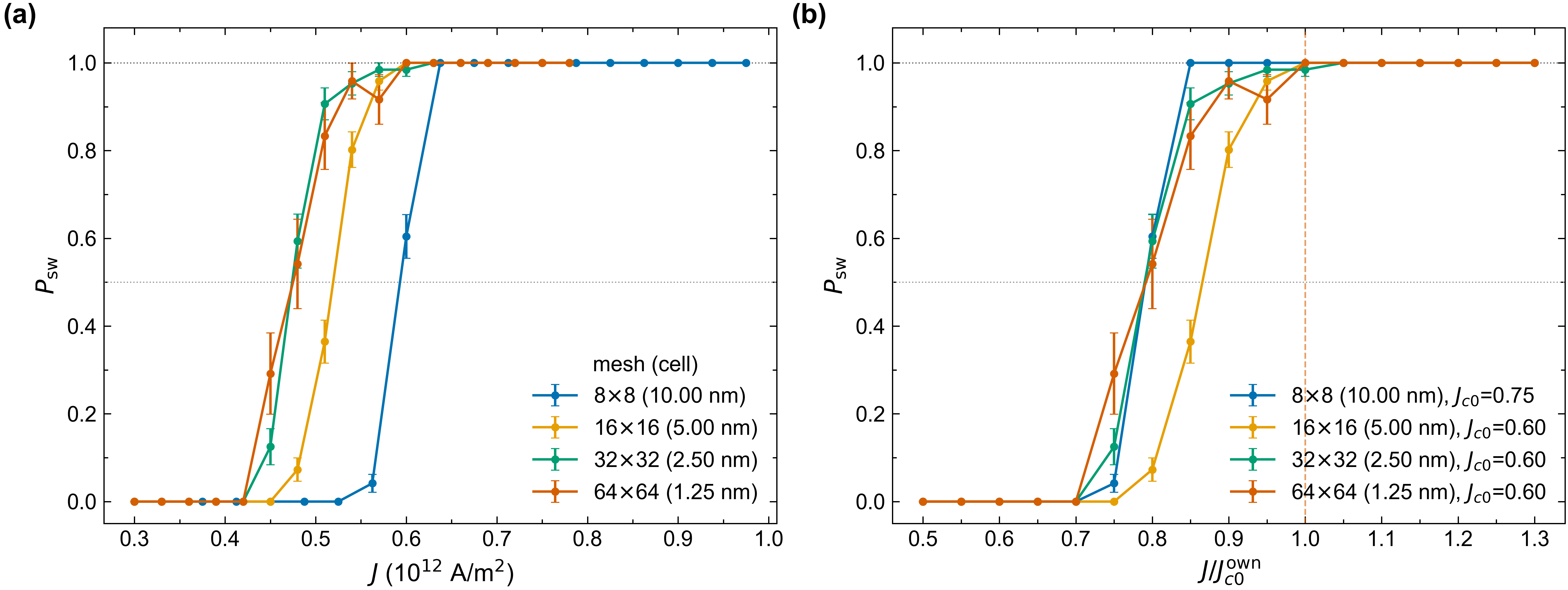}
\caption{\textbf{Mesh convergence of the STT switching threshold, enabled by
replica-batching throughput.} The same $80\times80\times1.5$\,nm CoFeB free
layer discretized on 8/16/32/64-cell meshes (10/5/2.5/1.25\,nm cells); the
absolute current is raised until $P_\mathrm{sw}=1$ (300\,K, 2\,ns; every trial
a batched replica). \textbf{(a)} The threshold converges once the cell size
drops below the exchange length $\ell_\mathrm{ex}=4.89$\,nm --- the 16/32/64
meshes coincide while the $8\times8$ mesh ($10$\,nm $>\ell_\mathrm{ex}$)
under-resolves the reversal and is biased high. \textbf{(b)} Normalizing each
curve by its own $J_{c0}$ nearly collapses them, showing the mesh effect is
almost entirely in the absolute threshold, not the transition shape.}
\label{fig:batchmesh}
\end{figure}


\begin{thebibliography}{99}
\footnotesize

\bibitem{landau1935} Landau, L. D. \& Lifshitz, E. M. On the theory of the
dispersion of magnetic permeability in ferromagnetic bodies.
\emph{Phys. Z. Sowjetunion} \textbf{8}, 153--169 (1935).

\bibitem{gilbert2004} Gilbert, T. L. A phenomenological theory of damping in
ferromagnetic materials. \emph{IEEE Trans. Magn.} \textbf{40}, 3443--3449
(2004).

\bibitem{brown1963book} Brown, W. F., Jr. \emph{Micromagnetics} (Interscience,
New York, 1963).

\bibitem{dieny2017} Dieny, B. \& Chshiev, M. Perpendicular magnetic
anisotropy at transition metal/oxide interfaces and applications.
\emph{Rev. Mod. Phys.} \textbf{89}, 025008 (2017).

\bibitem{ralph2008} Ralph, D. C. \& Stiles, M. D. Spin transfer torques.
\emph{J. Magn. Magn. Mater.} \textbf{320}, 1190--1216 (2008).

\bibitem{manchon2019} Manchon, A. \emph{et al.} Current-induced spin--orbit
torques in ferromagnetic and antiferromagnetic systems.
\emph{Rev. Mod. Phys.} \textbf{91}, 035004 (2019).

\bibitem{parkin2008} Parkin, S. S. P., Hayashi, M. \& Thomas, L. Magnetic
domain-wall racetrack memory. \emph{Science} \textbf{320}, 190--194 (2008).

\bibitem{fert2017} Fert, A., Reyren, N. \& Cros, V. Magnetic skyrmions:
advances in physics and potential applications. \emph{Nat. Rev. Mater.}
\textbf{2}, 17031 (2017).

\bibitem{muhlbauer2009} M\"uhlbauer, S. \emph{et al.} Skyrmion lattice in a
chiral magnet. \emph{Science} \textbf{323}, 915--919 (2009).

\bibitem{sampaio2013} Sampaio, J., Cros, V., Rohart, S., Thiaville, A. \&
Fert, A. Nucleation, stability and current-induced motion of isolated
magnetic skyrmions in nanostructures. \emph{Nat. Nanotechnol.} \textbf{8},
839--844 (2013).

\bibitem{fidler2000} Fidler, J. \& Schrefl, T. Micromagnetic modelling ---
the current state of the art. \emph{J. Phys. D: Appl. Phys.} \textbf{33},
R135--R156 (2000).

\bibitem{miltat2007} Miltat, J. E. \& Donahue, M. J. Numerical
micromagnetics: finite difference methods. In \emph{Handbook of Magnetism
and Advanced Magnetic Materials} Vol.~2 (Wiley, 2007).

\bibitem{abert2019} Abert, C. Micromagnetics and spintronics: models and
numerical methods. \emph{Eur. Phys. J. B} \textbf{92}, 120 (2019).

\bibitem{leliaert2019} Leliaert, J. \& Mulkers, J. Tomorrow's micromagnetic
simulations. \emph{J. Appl. Phys.} \textbf{125}, 180901 (2019).

\bibitem{oommf} Donahue, M. J. \& Porter, D. G. \emph{OOMMF User's Guide,
Version 1.0}. Interagency Report NISTIR 6376 (NIST, Gaithersburg, 1999).

\bibitem{vansteenkiste2014} Vansteenkiste, A. \emph{et al.} The design and
verification of MuMax3. \emph{AIP Adv.} \textbf{4}, 107133 (2014).

\bibitem{leliaert2018} Leliaert, J. \emph{et al.} Fast micromagnetic
simulations on GPU --- recent advances made with mumax$^3$.
\emph{J. Phys. D: Appl. Phys.} \textbf{51}, 123002 (2018).

\bibitem{mumaxplus} Moreels, L., Lateur, I., De Gusem, D. \emph{et al.}
mumax$^+$: extensible GPU-accelerated micromagnetics and beyond.
\emph{npj Comput. Mater.} \textbf{12}, 71 (2026).
\url{https://doi.org/10.1038/s41524-025-01893-y}

\bibitem{mumaxco} You, C.-Y. Optimization of MuMax3 by using Claude Code: a
CUDA-graph-based case study in AI-assisted performance engineering.
\emph{J. Magn.} \textbf{31}, 204 (2026).
\url{https://doi.org/10.4283/JMAG.2026.31.2.204}

\bibitem{magnumnp} Bruckner, F., Koraltan, S., Abert, C. \& Suess, D.
magnum.np: a PyTorch-based GPU-enhanced finite-difference micromagnetic
simulation framework for high-level development and inverse design.
\emph{Sci. Rep.} \textbf{13}, 12054 (2023).

\bibitem{boris} Lepadatu, S. Boris computational spintronics --- high
performance multi-mesh magnetic and spin transport modeling software.
\emph{J. Appl. Phys.} \textbf{128}, 243902 (2020).

\bibitem{fidimag} Bisotti, M.-A. \emph{et al.} Fidimag --- a finite
difference atomistic and micromagnetic simulation package. \emph{J. Open
Res. Softw.} \textbf{6}, 22 (2018).

\bibitem{mumag} Donahue, M. J. \& the \textmu MAG group. \emph{\textmu MAG
Standard Problems}. NIST/CTCMS,
\url{https://www.ctcms.nist.gov/~rdm/mumag.org.html}.

\bibitem{you2026vrqc} You, C.-Y. Variance-reduced trajectory unravelings for
GPU noisy quantum-circuit simulation: characterization and a Qiskit-Aer
integration gap. Preprint at \url{https://arxiv.org/abs/2607.17678} (2026).

\bibitem{vkfft} Tolmachev, D. VkFFT --- a performant, cross-platform and
open-source GPU FFT library. \emph{IEEE Access} \textbf{11}, 12039--12058
(2023).

\bibitem{kittel1948} Kittel, C. On the theory of ferromagnetic resonance
absorption. \emph{Phys. Rev.} \textbf{73}, 155--161 (1948).

\bibitem{rohart2013} Rohart, S. \& Thiaville, A. Skyrmion confinement in
ultrathin film nanostructures in the presence of Dzyaloshinskii--Moriya
interaction. \emph{Phys. Rev. B} \textbf{88}, 184422 (2013).

\bibitem{dzyaloshinskii1958} Dzyaloshinsky, I. A thermodynamic theory of
``weak'' ferromagnetism of antiferromagnetics. \emph{J. Phys. Chem. Solids}
\textbf{4}, 241--255 (1958).

\bibitem{moriya1960} Moriya, T. Anisotropic superexchange interaction and
weak ferromagnetism. \emph{Phys. Rev.} \textbf{120}, 91--98 (1960).

\bibitem{zhang2004} Zhang, S. \& Li, Z. Roles of nonequilibrium conduction
electrons on the magnetization dynamics of ferromagnets.
\emph{Phys. Rev. Lett.} \textbf{93}, 127204 (2004).

\bibitem{thiaville2012} Thiaville, A., Rohart, S., Ju\'e, \'E., Cros, V. \&
Fert, A. Dynamics of Dzyaloshinskii domain walls in ultrathin magnetic
films. \emph{Europhys. Lett.} \textbf{100}, 57002 (2012).

\bibitem{dormand1980} Dormand, J. R. \& Prince, P. J. A family of embedded
Runge--Kutta formulae. \emph{J. Comput. Appl. Math.} \textbf{6}, 19--26
(1980).

\bibitem{cudagraphs} NVIDIA Corporation. \emph{CUDA C++ Programming Guide}
--- CUDA Graphs; \emph{NVIDIA Blackwell Architecture Whitepaper} (2024).

\bibitem{brown1963} Brown, W. F., Jr. Thermal fluctuations of a
single-domain particle. \emph{Phys. Rev.} \textbf{130}, 1677--1686 (1963).

\bibitem{garcia1998} Garc\'ia-Palacios, J. L. \& L\'azaro, F. J. Langevin
dynamics study of the dynamical properties of small magnetic particles.
\emph{Phys. Rev. B} \textbf{58}, 14937--14958 (1998).

\bibitem{depondt2009} Depondt, P. \& Mertens, F. G. Spin dynamics simulations
of soft ferromagnetic nanoparticles at finite temperature.
\emph{J. Phys.: Condens. Matter} \textbf{21}, 336005 (2009).

\bibitem{sun2000} Sun, J. Z. Spin-current interaction with a monodomain
magnetic body: a model study. \emph{Phys. Rev. B} \textbf{62}, 570--578
(2000).

\bibitem{newell1993} Newell, A. J., Williams, W. \& Dunlop, D. J. A
generalization of the demagnetizing tensor for nonuniform magnetization.
\emph{J. Geophys. Res.} \textbf{98}(B6), 9551--9555 (1993).

\bibitem{slonczewski1996} Slonczewski, J. C. Current-driven excitation of
magnetic multilayers. \emph{J. Magn. Magn. Mater.} \textbf{159}, L1--L7
(1996).

\bibitem{berger1996} Berger, L. Emission of spin waves by a magnetic
multilayer traversed by a current. \emph{Phys. Rev. B} \textbf{54},
9353--9358 (1996).

\bibitem{fftw} Frigo, M. \& Johnson, S. G. The design and implementation of
FFTW3. \emph{Proc. IEEE} \textbf{93}, 216--231 (2005).
\end{thebibliography}
\end{document}